\documentclass[twoside,a4paper]{article}

\usepackage{cite}
\usepackage{amsmath}
\usepackage{amssymb}
\usepackage{graphicx}

\usepackage[T2A]{fontenc}
\usepackage[cp1251]{inputenc}

\usepackage{textcomp}

\usepackage{color}

\usepackage{cmpj2}
\usepackage{color}

\newcommand\tabcaption{\def\@captype{table}\caption}
\newcommand{\be}{\begin{eqnarray*}}
\newcommand{\ee}{\end{eqnarray*}}
\newcommand{\ibe}{\begin{eqnarray}}
\newcommand{\iee}{\end{eqnarray}}

\newcommand{\nb}{\nabla}

\issue{2016}{19}{3}{33001}
\doinumber{10.5488/CMP.19.33001}

\title[Numerical investigation of local defectiveness control of diblock copolymer patterns]%
{Numerical investigation of local defectiveness control of diblock copolymer patterns}
\author[D. Jeong, Y. Choi, J. Kim]{D. Jeong,
        Y. Choi, J. Kim\thanks{Corresponding author, E-mail: cfdkim@korea.ac.kr.
}}
\address{Department of Mathematics, Korea University, Seoul 136-713, Republic of Korea
}

\authorcopyright{D. Jeong, Y. Choi, J. Kim, 2016}
\date{Received October 21, 2015, in final form December 22, 2015}

\begin{document}

\maketitle

\begin{abstract}
We numerically investigate local defectiveness control of
self-assembled diblock copolymer patterns through appropriate
substrate design. We use a nonlocal Cahn-Hilliard (CH) equation for
the phase separation dynamics of diblock copolymers. We
discretize the nonlocal CH equation by an unconditionally stable
finite difference scheme on a tapered trench design and, in
particular, we use Dirichlet, Neumann, and periodic boundary
conditions. The value at the Dirichlet boundary comes from an
energy-minimizing equilibrium lamellar profile. We solve the
resulting discrete equations using a Gauss-Seidel iterative method. We
perform various numerical experiments such as effects of channel
width, channel length, and angle on the phase separation dynamics.
The simulation results are consistent with the previous experimental
observations.
\keywords diblock copolymer, nonlocal Cahn-Hilliard equation, local
defectivity control
\pacs 02.60.Cb, 02.60.Lj, 02.70.Bf, 02.70.Pt
\end{abstract}

\section{Introduction}

A diblock copolymer is a linear chain consisting of two blocks of
different types of monomers bonded covalently to each other. The two
blocks are mixed above the critical temperature; however, the
copolymer melt undergoes phase separation below the critical
temperature due to the incompatibility of different blocks
\cite{DJSLYCJK2015}. As a result of phase separation, periodic
structures including lamellae \cite{CPW,FUB,JSL,LCKY,SW,YS}, spheres
\cite{CPW,FUB,GTY,TSDMK2010,LZT,YT,ZDY}, cylinders
\cite{CPW,FUB,SW,LZT,JH,PLX}, hexagons
\cite{CPW,FUB,YS,LZT,JH,PLX,LJC,MS2015,POM}, and gyroids
\cite{CPW,FUB,LZT} are observed in a mesoscopic-scale domain.

\begin{figure}[!h]
\centering
\begin{minipage}{0.24\linewidth}
\begin{center}
\includegraphics[width=1.32in]{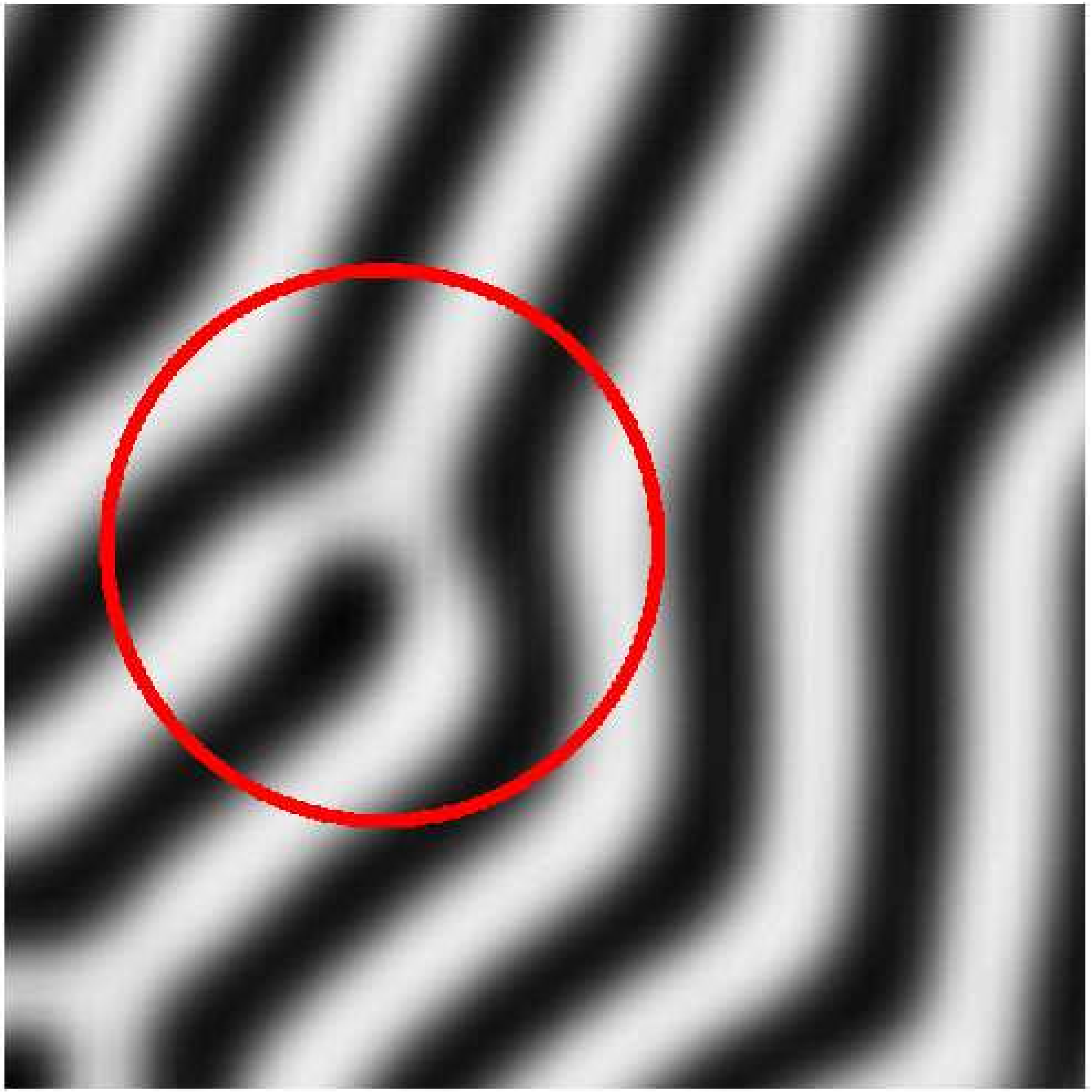}
\end{center}
\end{minipage}
\begin{minipage}{0.24\linewidth}
\begin{center}
\includegraphics[width=1.32in]{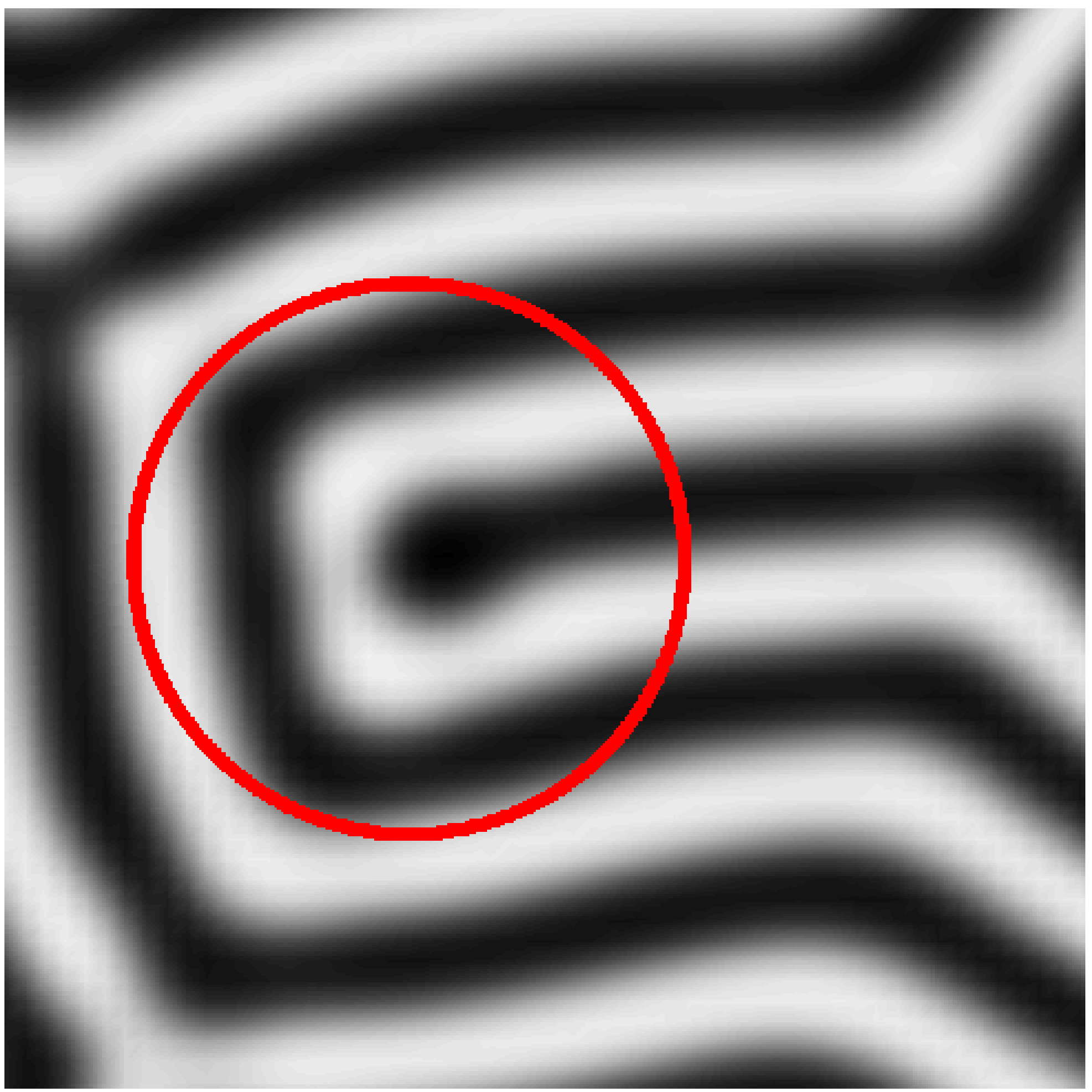}
\end{center}
\end{minipage}
\begin{minipage}{0.24\linewidth}
\begin{center}
\includegraphics[width=1.32in]{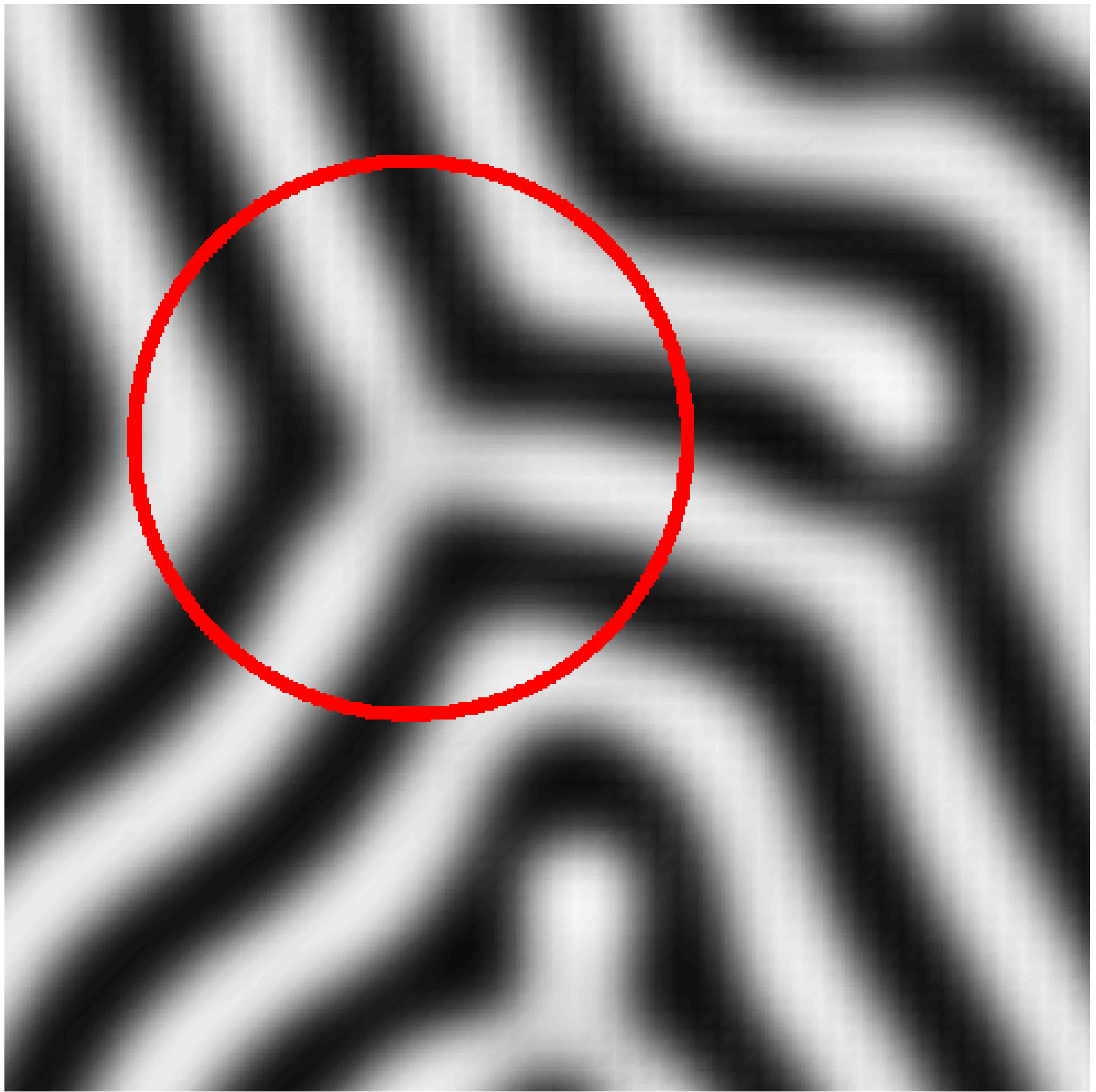}
\end{center}
\end{minipage}
\begin{minipage}{0.24\linewidth}
\begin{center}
\includegraphics[width=1.32in]{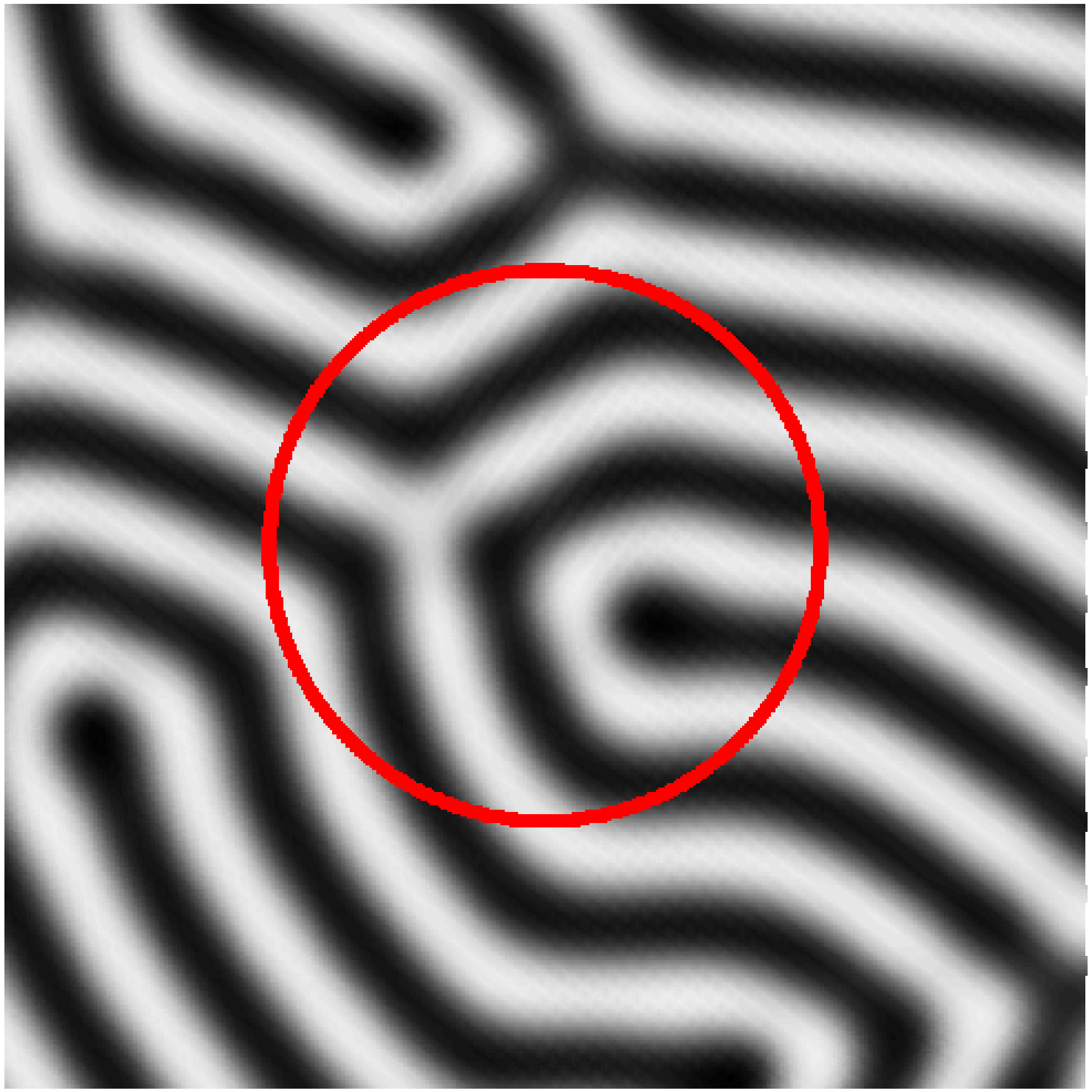}
\end{center}
\end{minipage}
\caption{Examples of local defects.} \label{defact}
\end{figure}

In recent years, self-assembly of block copolymer has come out as a
promising patterning tool to overcome the scaling limits in
nano-lithography and generate suboptical lithographic patterns
\cite{KLDC}. However, one of the problems is the lack of complete
pattern orientation due to a high density of defects
\cite{RRZS}. In figure~\ref{defact}, we can observe various
examples of local defect in the block copolymer. Therefore, it is
very important to control the local defects of self-assembled
polymer patterns with the application of these materials. As the efforts
to rectify this, many researches and techniques such as electric
fields \cite{MLUE}, flow \cite{PARC}, shear application
\cite{AWAD,H2015,GZWWSP2015}, thermal treatment \cite{SHJ2011},
chemically pre-patterned surface (chemoepitaxy) \cite{KSSF,LRHC},
and topographical confinement (graphoepitaxy) \cite{SYK} have been
carried out to reduce the defect density in specific pattern-forming
block copolymer thin films. Among the controlling method, authors in
\cite{RRZS} proposed an appropriate substrate design and achieved a
defect-free pattern formation. In this paper, we focus on
numerically realizing the situation presented in \cite{RRZS} and we
describe in detail the numerical method which is used in the
numerical simulations.

We use the mathematical model proposed by Ohta and Kawasaki
\cite{OK}. Let $\phi$ be {the difference of the local
volume fraction of $A$ and $B$ monomers.} Then, the nonlocal
Cahn-Hilliard (CH) equation in a two-dimensional domain is
 \ibe
    \frac{\partial \phi({\bf x},t)}{\partial t} &=& \Delta \mu({\bf x},t)
    - \alpha \left[ \phi({\bf x},t) - \bar\phi \right], \label{govern1} \\
    \mu({\bf x},t) &=&   F'\big(\phi({\bf x},t)\big) - \epsilon^2 \Delta \phi({\bf x},t) \label{govern2},
 \iee
where ${\bf x}=(x,y)$ and $t$ are the spatial and temporal
variables, respectively. $F(\phi) = 0.25 (\phi^2-1)^2$ is the
Helmholtz free energy, $\epsilon$ is the gradient energy
coefficient, $\alpha$ is inversely proportional to the square of the
total chain length of the copolymer, and $\bar\phi = \int_\Omega
\phi({\bf x},0) \rd{\bf x} / |\Omega|$ is the average concentration
over the domain $\Omega$ \cite{NO}.

{ In equation (\ref{govern1}), $\alpha [ \phi({\bf x},t)
- \bar\phi ]$ term indicates the long-range interaction and
plays an important part in pattern formation. If $\alpha = 0$, then
equations (\ref{govern1}) and (\ref{govern2}) describe the process of the
reduction in the total interfacial energy of a microstructure as the
classical CH equation.

 The total system energy is given as
 \ibe
    \mathcal{E}(\phi) = \int_\Omega \left[ F(\phi) +
    \frac{\epsilon^2}{2} \left| \nb \phi \right|^2 \right] \rd{\bf x}
     + \frac{\alpha}{2} \int_\Omega \int_\Omega G({\bf x}-{\bf y})\left[\phi({\bf x})-\bar \phi\right]\left[\phi({\bf y})-\bar \phi\right]
    \rd{\bf y} \rd{\bf x}\, ,\label{origene}
 \iee
where $G$ is the Green's function of $-\Delta$ in $\Omega$ with
periodic boundary conditions, i.e., $-\Delta G({\bf x})=\delta ({\bf
x}) $. Then, the evolving equations (\ref{govern1}) and (\ref{govern2})
can be derived using the $H^{-1}$ gradient flow for the free energy
(\ref{origene}), and equation (\ref{origene}) can be rewritten as
 \be
    \mathcal{E}(\phi) = \int_\Omega \left[ F(\phi) +
    \frac{\epsilon^2}{2} \left| \nb \phi \right|^2 \right] \rd{\bf x} +
    \frac{\alpha}{2} \int_\Omega \left| \nb \psi \right|^2 \rd{\bf x}\, ,
 \ee
where $\psi$ satisfies $-\Delta \psi = \phi - \bar \phi$ with
periodic boundary conditions \cite{CPW}. }

\begin{figure}[!b]
\begin{center}
\includegraphics[width=5.0in]{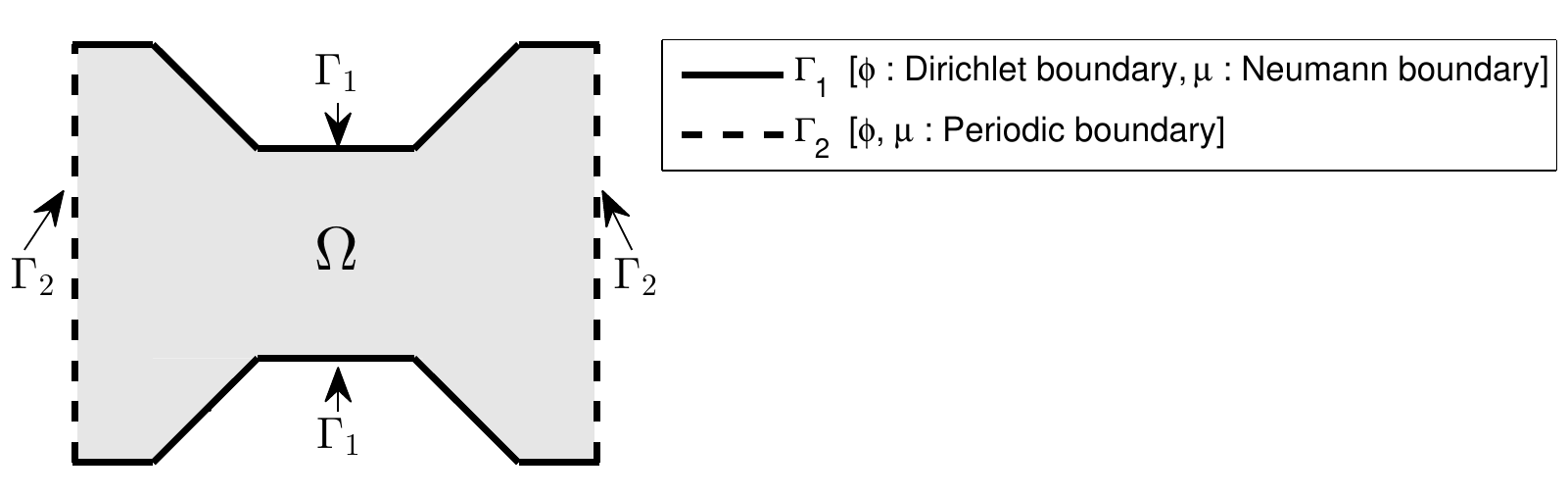}
\end{center}
\caption{Illustration of the physical domain ($\Omega$) with
boundaries $\Gamma_1$ and $\Gamma_2$.}
\label{schematic_channel_info.eps00}
\end{figure}

Now, we will solve equations (\ref{govern1}) and (\ref{govern2}) on a
trench domain. Figure \ref{schematic_channel_info.eps00} represents
the physical domain ($\Omega$) and boundaries
($\Gamma_1$, $\Gamma_2$). On $\Gamma_1$, Dirichlet boundary condition
for $\phi$ and homogeneous Neumann boundary condition for $\mu$ are
used. On $\Gamma_2$, the periodic boundary condition for both
$\phi$ and $\mu$ is used.

The rest of this paper is organized as follows. In section~\ref{nmethod}, we describe the numerical method and solution. In
section~\ref{nex}, we present several numerical experiments.
Conclusions are summarized in section~\ref{Conc}.

\section{Numerical method} \label{nmethod}

\subsection{Discretization of domain}

\begin{figure}[htbp]
\begin{center}
\includegraphics[width=5.0in]{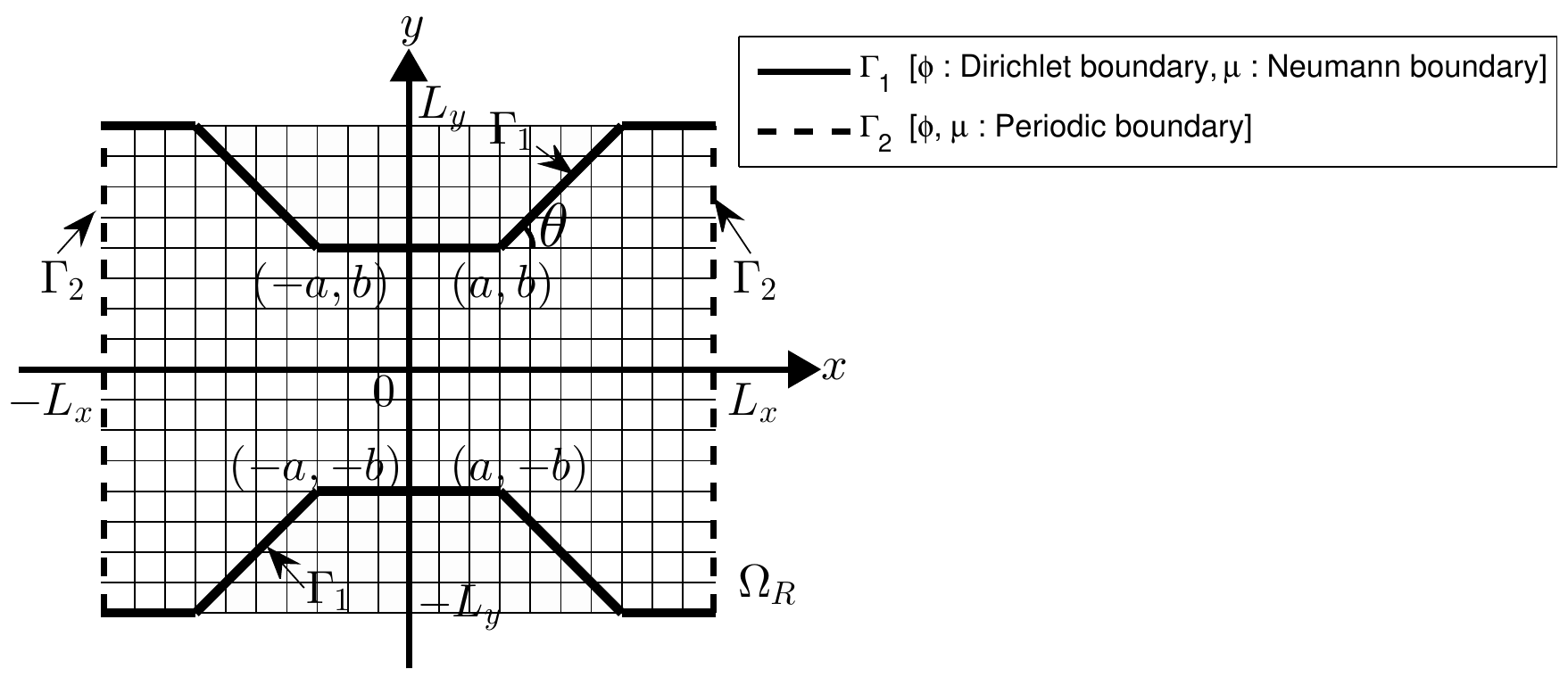}
\end{center}
\caption{Illustration of the parameters over the whole domain
$\Omega_\text{R} = (-L_x,L_x)\times(-L_y,L_y)$. $\Gamma_1$ and $\Gamma_2$
are boundary of the computational domain which is determined from
$\theta$. Trench walls are defined with symmetric points
$(-a,b)$, $(a,b)$, $(-a,-b)$, and $(a,-b)$.}
\label{schematic_channel_info}
\end{figure}

 First, assume that we have a domain $\Omega$ as shown
in figure \ref{schematic_channel_info.eps00}. The domain $\Omega$ is
defined by the angle $\theta$, reference values $a$ and $b$ for the
trench wall as represented in figure \ref{schematic_channel_info}.
Here, the trench walls are determined with symmetric points
$(-a,b)$, $(a,b)$, $(-a,-b),$ and $(a,-b)$. Then, we cover the domain
$\Omega$ by a rectangular domain $\Omega_\text{R} = (-L_x,L_x)\times
(-L_y,L_y)$ with a Cartesian grid of mesh size $h$.

Now, we discretize the rectangular domain $\Omega_\text{R}$ with the uniform
mesh size $h=2L_x/N_x=2L_y/N_y$ in both $x$- and $y$-directions.
Here, $N_x$ and $N_y$ are the number of grid points in $x$- and $y$-directions, respectively. We denote cell-corner points as $(x_i,y_j)
= (h_i, h_j)$ for $i=0, \ldots, N_x$ and  $j=0, \ldots, N_y$. Let
$\phi^n_{ij}$ and $\mu^n_{ij}$ be approximations of
$\phi(x_i,y_j,t_n)$ and $\mu(x_i,y_j,t_n)$, respectively, where $t_n
= n \Delta t$ and $\Delta t$ is the temporal step size.

\subsection{Numerical solution}

\begin{figure}[!t]
\begin{center}
\includegraphics[width=5.0in]{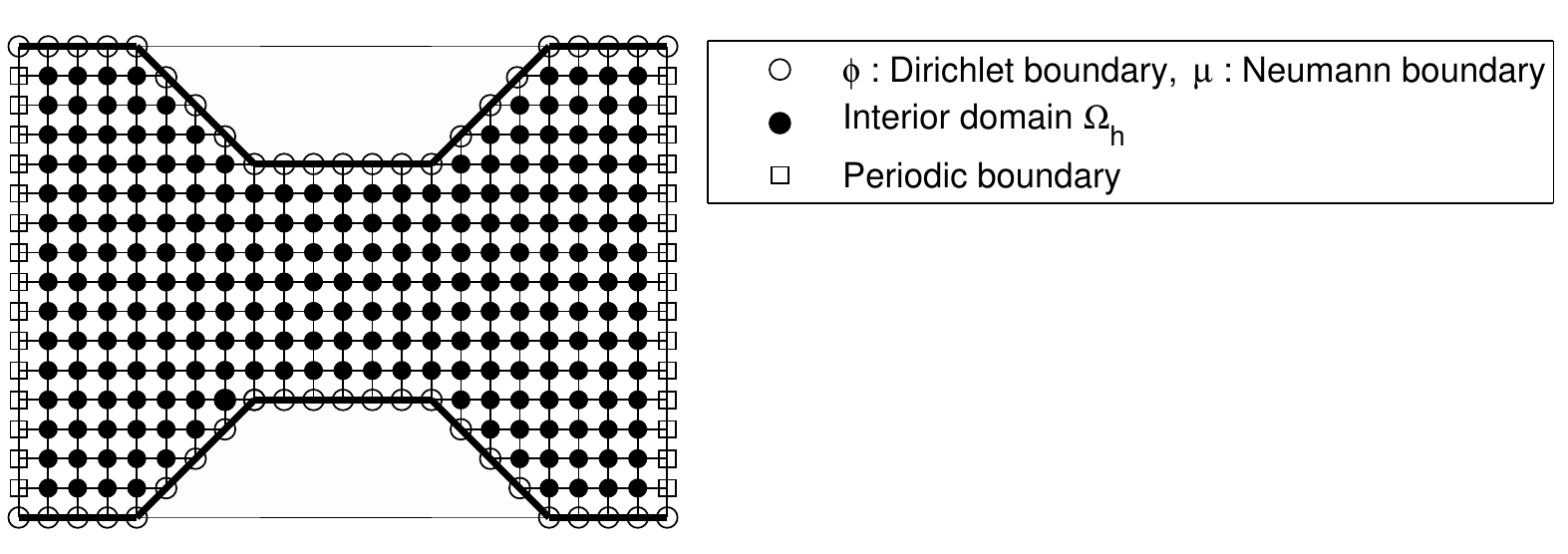}
\end{center}
\caption{Inner grid points ($\bullet$) which are on the
computational domain $\Omega_h$, Dirichlet ($\phi$) and homogeneous
Neumann ($\mu$) boundary points ($\circ$), and periodic boundary
points ($\Box$).} \label{schematic_channel}
\end{figure}

 In this paper, we apply a non-linearly stabilized
splitting scheme \cite{E1998} to the nonlocal CH equations
(\ref{govern1}) and (\ref{govern2}) as follows:
 \ibe
\frac{\phi^{n+1}_{ij}-\phi^n_{ij}}{\Delta t}
&=& \Delta_h \mu^{n+1}_{ij}  - \alpha \left( \phi^{n+1}_{ij} - {\bar \phi} \right), \label{CH_sp2} \\
\mu^{n+1}_{ij} &=&  \left( \phi^{n+1}_{ij}\right)^3  -\phi^n_{ij}- \epsilon^2
\Delta_h \phi^{n+1}_{ij}, \label{CH_sp3}
 \iee
where $\bar \phi = \sum_{{\bf x}_{ij} \in \Omega_h} \phi_{ij}^0 \big/
\sum_{{\bf x}_{ij} \in \Omega_h } 1$. Here, $\Omega_h$ is the
computational domain which is represented by marked circle in figure
\ref{schematic_channel}.

 To solve equations (\ref{CH_sp2}) and  (\ref{CH_sp3}), we use
the Gauss--Seidel iterative method. Given solution $\phi_{ij}^n$,
let $\phi_{ij}^{n+1,0}=\phi_{ij}^n$ be an initial guess. For each $m
\geqslant 0$, we generate the updated solution $\phi_{ij}^{n+1,m+1}$ and
$\mu_{ij}^{n+1,m+1}$ from $\phi_{ij}^{n+1,m}$ and $\mu_{ij}^{n+1,m}$
by
\ibe
 \left( \frac{1}{\Delta t} + \alpha
\right) \phi_{ij}^{n+1,m+1} + \frac{4}{h^2} \mu_{ij}^{n+1,m+1}
= \frac{\phi_{ij}^n}{\Delta t} + \alpha \bar{\phi}
 + \frac{\mu_{i-1,j}^{n+1,m+1}+\mu_{i+1,j}^{n+1,m}+\mu_{i,j-1}^{n+1,m+1}+\mu_{i,j+1}^{n+1,m}}{h^2}\, ,
\iee
\begin{align}
\left[ -\frac{4 \epsilon^2}{h^2} -3\left(\phi_{ij}^{n+1,m}\right)^2 \right]
\phi_{ij}^{n+1,m+1} + \mu_{ij}^{n+1,m+1}
&= -\phi_{ij}^n -2\left(\phi_{ij}^{n+1,m}\right)^3\nonumber\\
&\quad-\epsilon^2
\frac{\phi_{i-1,j}^{n+1,m+1}+\phi_{i+1,j}^{n+1,m}+\phi_{i,j-1}^{n+1,m+1}+\phi_{i,j+1}^{n+1,m}}{h^2}\, .
 \end{align}
We continue the above iterations until $l_2$-norm error between two
successive approximations of $\phi$ is less than a given tolerance
$\mathrm{tol}$, that is,
 \be
 \left\|\phi^{n+1,m+1} - \phi^{n+1,m}\right\|_2 < \mathrm{tol}.
 \ee

\subsection{Boundary conditions}

 For a numerical solution, we consider three
different conditions at each boundary as follows:
 \begin{itemize}
 \item $\phi_{ij} = \|\phi^\text{eq}\|_{\infty}$  for  ${\bf x}_{ij} \in \Gamma_1$.
 \item $ \nabla_h \mu_{ij} = 0$  for  ${\bf x}_{ij} \in \Gamma_1$.
 \item $\phi_{0j} = \phi_{N_x+1,j}$ and $\mu_{0j} = \mu_{N_x+1,j}$  for  $j = 1, \ldots, N_y+1$.
 \end{itemize}
Here, $\|\phi^\text{eq}\|_{\infty}$ represents the maximum value of
numerical solution at equilibrium state. In subsection \ref{opt}, we
will describe more details for $\|\phi^\text{eq}\|_{\infty}$.

\begin{figure}[!b]
\begin{minipage}{0.49\linewidth}
\begin{center}
\includegraphics[width=2.25in]{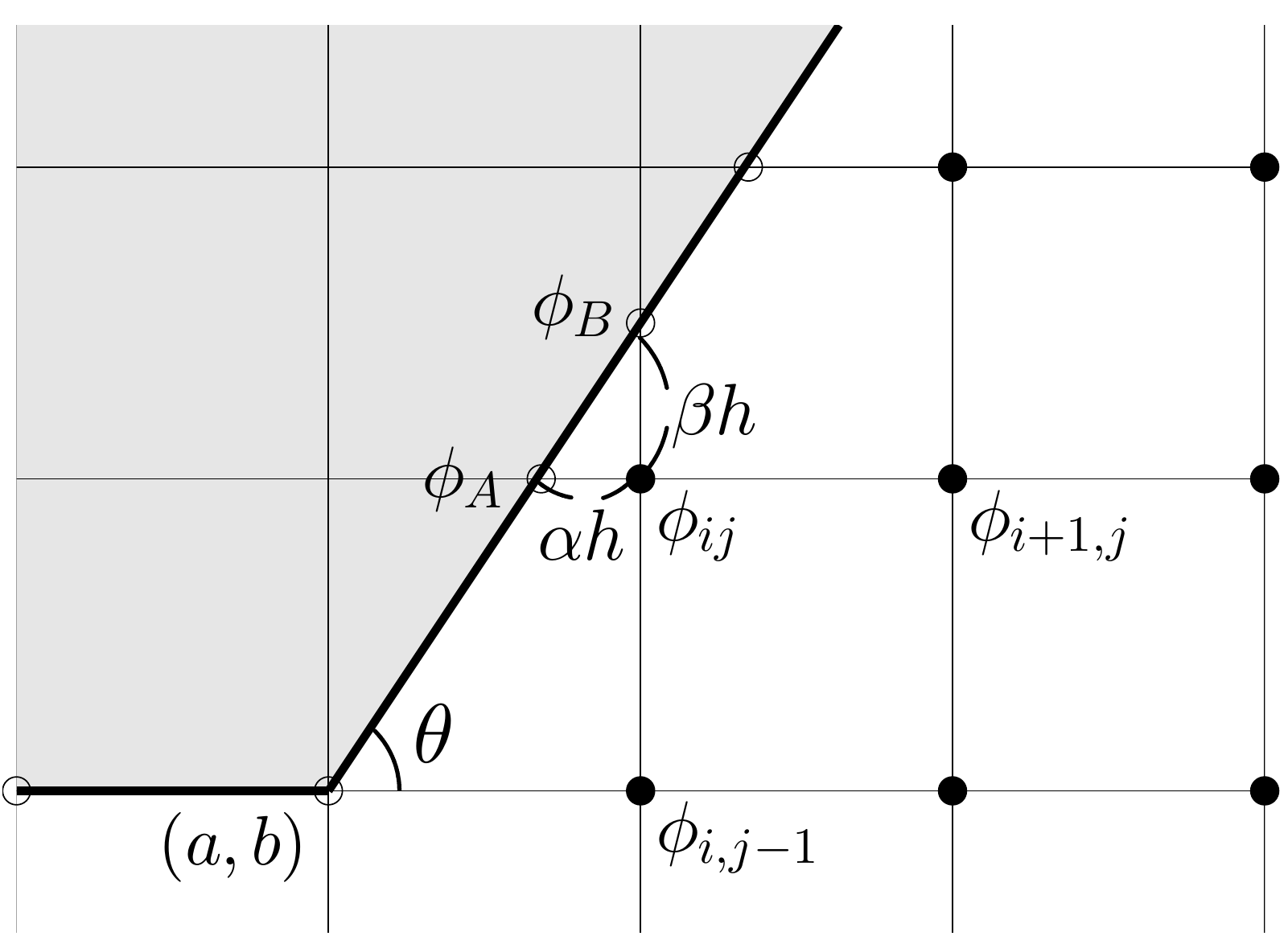}\\
(a)
\end{center}
\end{minipage}
\begin{minipage}{0.49\linewidth}
\begin{center}
\includegraphics[width=2.25in]{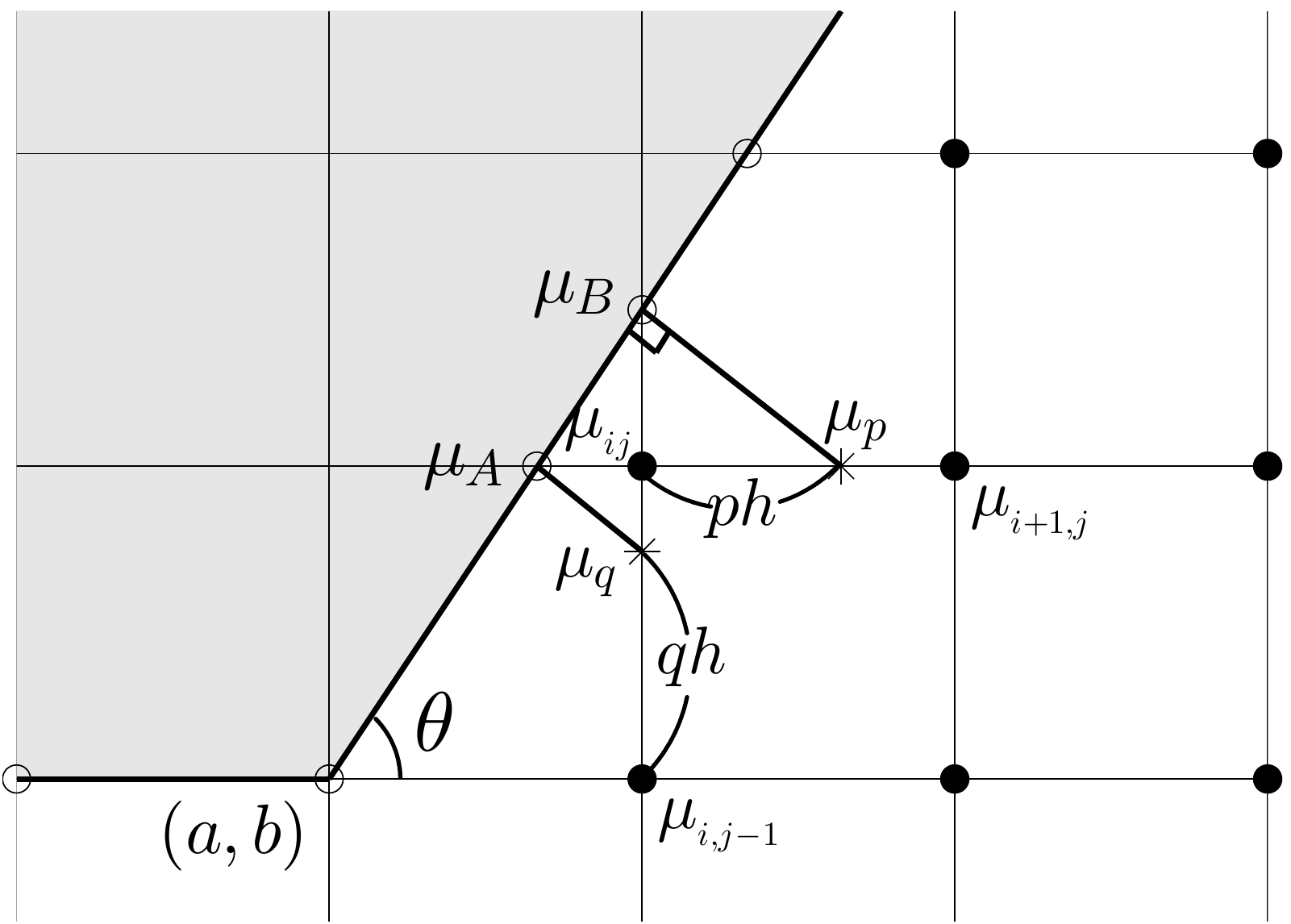}\\
(b)
\end{center}
\end{minipage}
\caption{(a) Dirichlet condition and (b) Neumann condition on curved
boundary.} \label{schematic_interp_mu}
\label{schematic_interp_phi}
\end{figure}

Near the boundaries, we should use some special formulae. For
example, let us consider the position $(x_i,y_j)$ in figure
\ref{schematic_interp_phi}. By the Dirichlet boundary condition,
we already know the value at $A$ and $B$. We define $\Delta_{xx}^\text{D}$
and $\Delta_{yy}^\text{D}$ as the discrete second derivatives near the
boundary as follows:
 \ibe
\Delta_{xx}^\text{D} \phi_{ij} &=& \left( \frac{\phi_{i+1,j}-\phi_{ij}}{h} -
\frac{\phi_{ij}-\phi_A}{\alpha h} \right)  \left( \frac{\alpha h
+h}{2}  \right)^{-1}, \\ \Delta_{yy}^\text{D} \phi_{ij} &=& \left(
\frac{\phi_B-\phi_{ij}}{\beta h} - \frac{\phi_{ij}-\phi_{i,j-1}}{h}
\right) \left( \frac{\beta h +h}{2} \right)^{-1},
 \iee
where $0< \alpha,~\beta <1$, and $\phi_A=\phi_B=
\|\phi^\text{eq}\|_{\infty}$. Therefore, the discrete Laplacian operator
near the boundary with Dirichlet condition is defined as $ \Delta_h^\text{D}
\phi^{n+1}_{ij} = \Delta_{xx}^\text{D} \phi_{ij}^{n+1}+ \Delta_{yy}^\text{D}
\phi_{ij}^{n+1}$. For other points, the discrete Laplacian is
similarly defined. We also define the discrete Laplacian operator
near the boundary with Neumann boundary condition as $ \Delta_h^\text{N}
\mu^{n+1}_{ij} = \Delta_{xx}^\text{N} \mu_{ij}^{n+1}+ \Delta_{yy}^\text{N}
\mu_{ij}^{n+1}$. Here,
\ibe
\Delta_{xx}^\text{N} \mu_{ij} &=& \Bigg(\frac{\mu_{i+1,j}-\mu_{ij}}{h} - \frac{\mu_{ij}-\mu_q}{\alpha h} \Bigg)
  \left( \frac{\alpha h
+h}{2}  \right)^{-1}, \\
\Delta_{yy}^\text{N} \mu_{ij} &=& \left( \frac{\mu_p-\mu_{ij}}{\beta h} -
\frac{\mu_{ij}-\mu_{i,j-1}}{h} \right)  \left( \frac{\beta h +h}{2}
\right)^{-1},
\iee
where $\alpha$ and $\beta$ are defined as in figure
\ref{schematic_interp_phi}~(a). $\mu_p$ and $\mu_q$ are obtained
by using a linear interpolation, $ \mu_p = p\mu_{i+1,j}+(1-p)
\mu_{ij} \mbox{ and } \mu_q = q\mu_{ij}+(1-q) \mu_{i,j-1}$ [see figure
\ref{schematic_interp_mu}~(b)].

\subsection{Optimal wavelength having minimum discrete total energy}
\label{opt}

\begin{figure}[!b]
\begin{minipage}{0.5\linewidth}
\begin{center}
\includegraphics[width=0.95\textwidth]{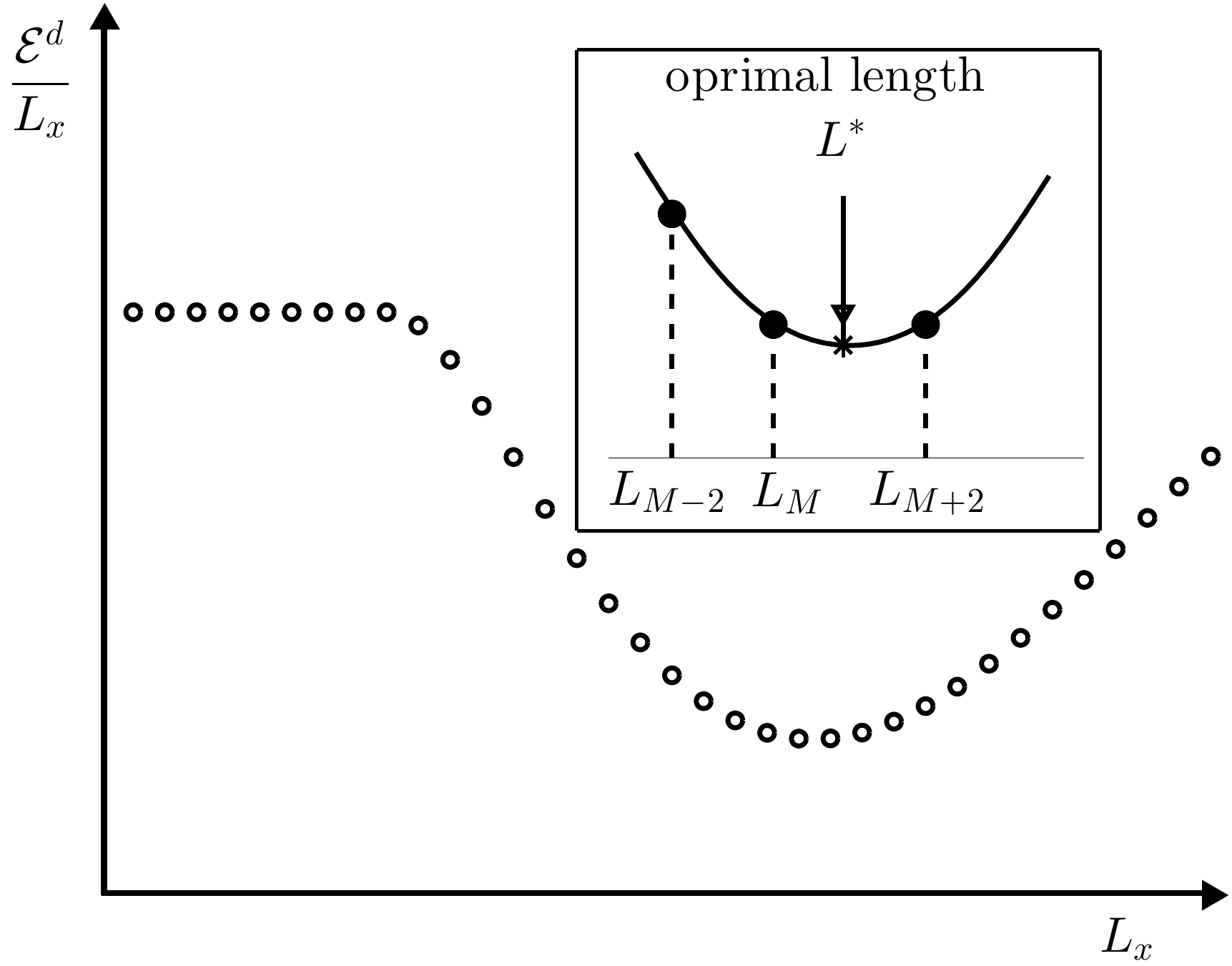} \\
\end{center}
\end{minipage}
\begin{minipage}{0.5\linewidth}
\begin{center}
\vspace{-5mm}
\includegraphics[width=0.95\textwidth]{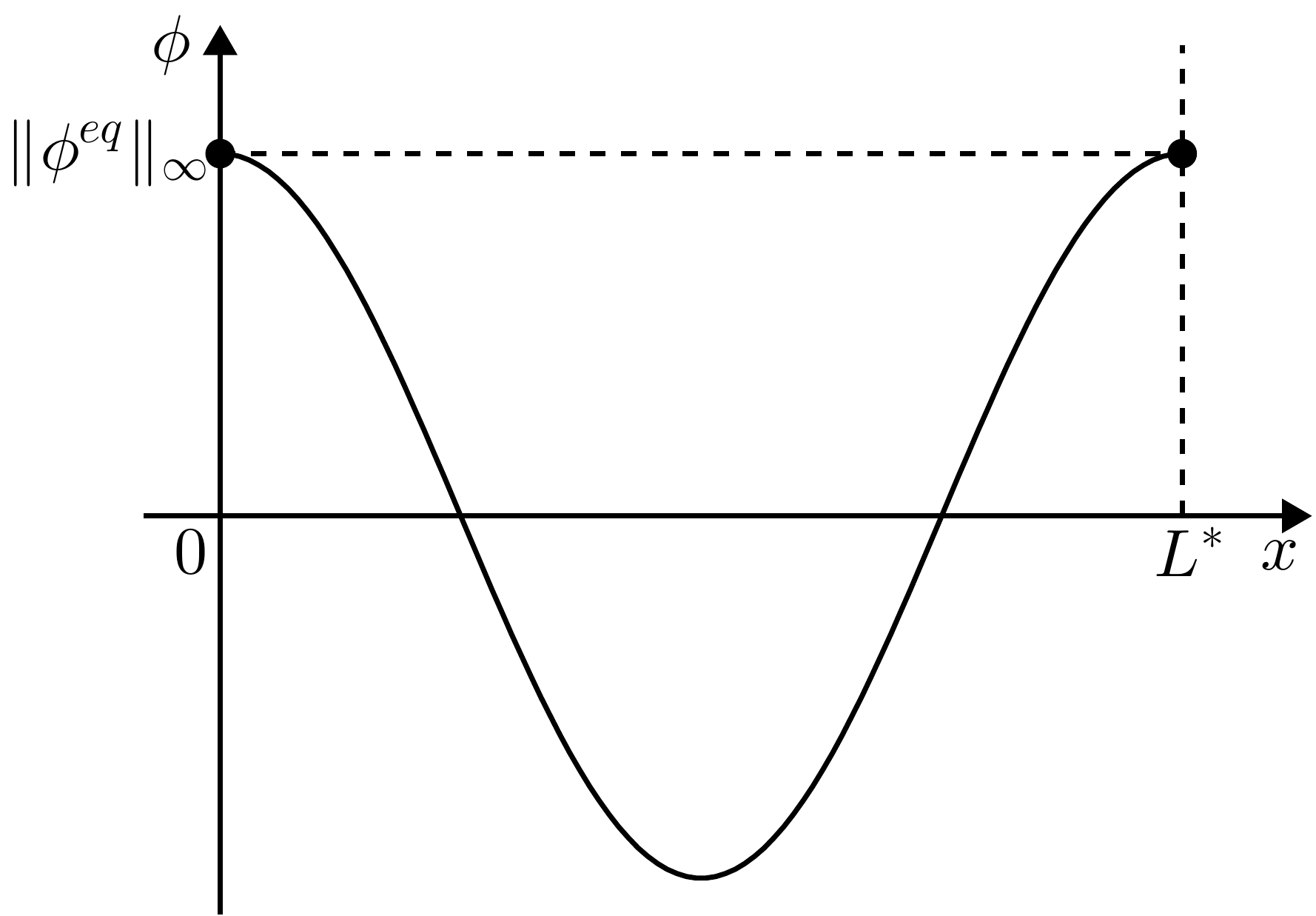} \\
\end{center}
\end{minipage}
\centerline{(a) \hspace{0.5\textwidth} (b)}
\caption{(a) Schematic of algorithm to search for the optimal length
$L^*$. Here, $L_{M-2} = (M-2) h_x\,$, $L_{M} = M h_x\,$, and $L_{M+2} =
(M+2) h_x\,$. (b) Illustration of maximum value $\| \phi^\text{eq}
\|_\infty$ of equilibrium wave.} \label{schematic_optimal}
\end{figure}

 We describe an algorithm for finding the
total energy-minimizing wavelength \cite{JSL,DJSLYCJK2015}. We
define the optimal wavelength $L^*$ as the period of the hexagonal
lattice that has the lowest energy. In other words, $L^*$ means the
smallest length having the global minimum of the domain-scaled
discrete total energy. To calculate $L^*$, we solve equations
(\ref{govern1}) and (\ref{govern2}) until a numerical equilibrium
state is reached with the given values of $h_x\,$, $\Delta t$, $\epsilon$,
and $\alpha$. The initial condition is $\phi(x,0)=0.1 \cos(2\pi x/
L_x) $ in $\Omega=(0, L_x)$, where $L_x$ starts at $2h_x$ and
increases in steps of $2h_x\,$. Let $M$ be the smallest even integer
such that the domain-scaled total energy $\mathcal{E}^\text{d}/L_x$ is
minimized. Construct the quadratic polynomial passing the three
points $\left((M-2)h_x\, ,\;\mathcal{E}^\text{d}/[(M-2)h_x]\right)$, $\left(Mh_x\, ,\;
\mathcal{E}^\text{d}/(Mh_x)\right)$, and $\left((M+2)h_x\, ,\; \mathcal{E}^\text{d}/[(M+2)h_x]\right)$;
then, define the optimal length $L^*$ as the critical point of the
polynomial [see figure \ref{schematic_optimal}~(a)]. For more details,
see references~\cite{DJSLYCJK2015,JSL}.

We define the numerical equilibrium state as that in which the
consecutive error is not larger than the prescribed tolerance, that
is, $\max_{1 \leqslant i \leqslant N_x}(|\phi_i^{k+1}-\phi_i^k|)/\Delta t \leqslant
1.0 \times 10^{-6}$. The maximum value of equilibrium wave is defined as $\|
\phi^\text{eq} \|_\infty = \max_{1\leqslant i \leqslant N_x} |\phi_i^\text{eq}|$ in figure
\ref{schematic_optimal}~(b).

{We replace the Dirichlet problem solution with $\|
\phi^\text{eq} \|_\infty$ in this paper.}


\section{Numerical experiments} \label{nex}

 In this section, we perform a number of numerical
tests. Throughout the numerical experiments, unless otherwise
specified, we use $\epsilon=1/(20\sqrt{2})$, $\alpha=100$,
$L^*=0.375$, $h=L^*/10$, $\Delta t=0.1h$, $\| \phi^\text{eq}\|_\infty =
0.6134$, and
 $\theta=\pi/4$. We examine the evolution of a random perturbation
about the average concentration $\bar{\phi} = 0$ on simple rectangle
domain $\Omega_\text{R} = (-25L^*,25L^*)\times(-15L^*, 15L^*)$ with
$N_x=500$, $N_y=300$. The initial condition is set to $\phi(x,y,0) =
{\bar \phi} + 0.01\;\textrm{rand}(x,y)$. Here, $\textrm{rand}(x,y)$ is
a random number between $-1$ and $1$. Also, we use $\mathrm{tol} = 10^{-4}$
for stopping criterion of the Gauss-Seidel iteration.

\subsection{Discrete total energy}

\begin{figure}[!h]
\begin{center}
\includegraphics[width=3.5in]{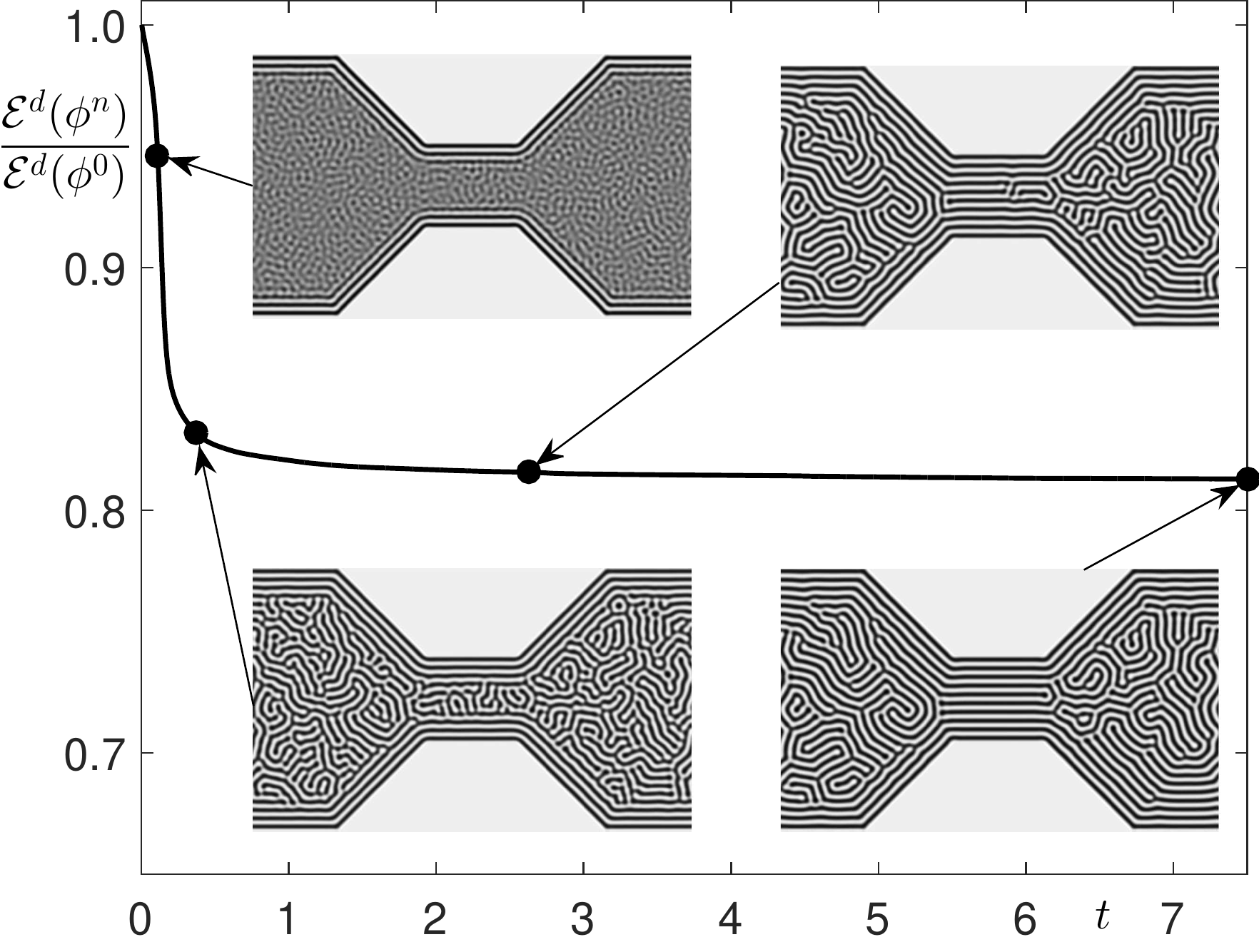}
\end{center}
\caption{{Time evolution of the normalized discrete
total energy $\mathcal{E}^\text{d} (\phi^n)/\mathcal{E}^\text{d}(\phi^0)$. Here,
the small figures indicate the concentration field $\phi$ at times
$t = 30\Delta t$, $100\Delta t$, $700\Delta t$, $2000\Delta t$,
respectively.}} \label{fig:energy}
\end{figure}

{We first define the discrete total energy as
 \be
 &&\mathcal E^\text{d}(\phi^n) = \sum_{i=1}^{N_x}\sum_{j=1}^{N_y}
\left\{ h^2F(\phi_{ij}^n)  + \frac{\epsilon^2}{2}
\left[\left(\phi_{i+1,j}^n-\phi_{ij}^n\right)^2 +\left(\phi_{i,j+1}^n-\phi_{ij}^n\right)^2 \right] \right.\\
 &&\left.\quad\phantom{\frac{\epsilon^2}{2}}  \qquad \qquad \qquad + \frac{\alpha}{2}  \left[
 \left(\psi_{i+1,j}^n-\psi_{ij}^n\right)^2+\left(\psi_{i,j+1}^n-\psi_{ij}^n\right)^2
 \right] \right\}.
 \ee
Note that $\psi$ satisfies $-\Delta \psi = \phi - \bar{\phi}$ with
periodic boundary conditions \cite{CPW}.

Figure \ref{fig:energy} shows the temporal evolution of the
normalized discrete total energy $\mathcal{E}^\text{d}
(\phi^n)/\mathcal{E}^\text{d}(\phi^0)$. In figure \ref{fig:energy}, we
can see that the normalized discrete total energy (which is denoted by
the solid line) is nonincreasing as time proceeds. Moreover, the
four small figures represent the numerical solution at times $t =
30\Delta t$, $100\Delta t$, $700\Delta t$, $2000\Delta t$, respectively.
}

\subsection{The effect of channel width}

\begin{figure}[!t]
\vspace{2mm}
\begin{minipage}{0.03\linewidth}
\centering (a)
\end{minipage}
\begin{minipage}{0.31\linewidth}
\begin{center}
\includegraphics[width=1.6in]{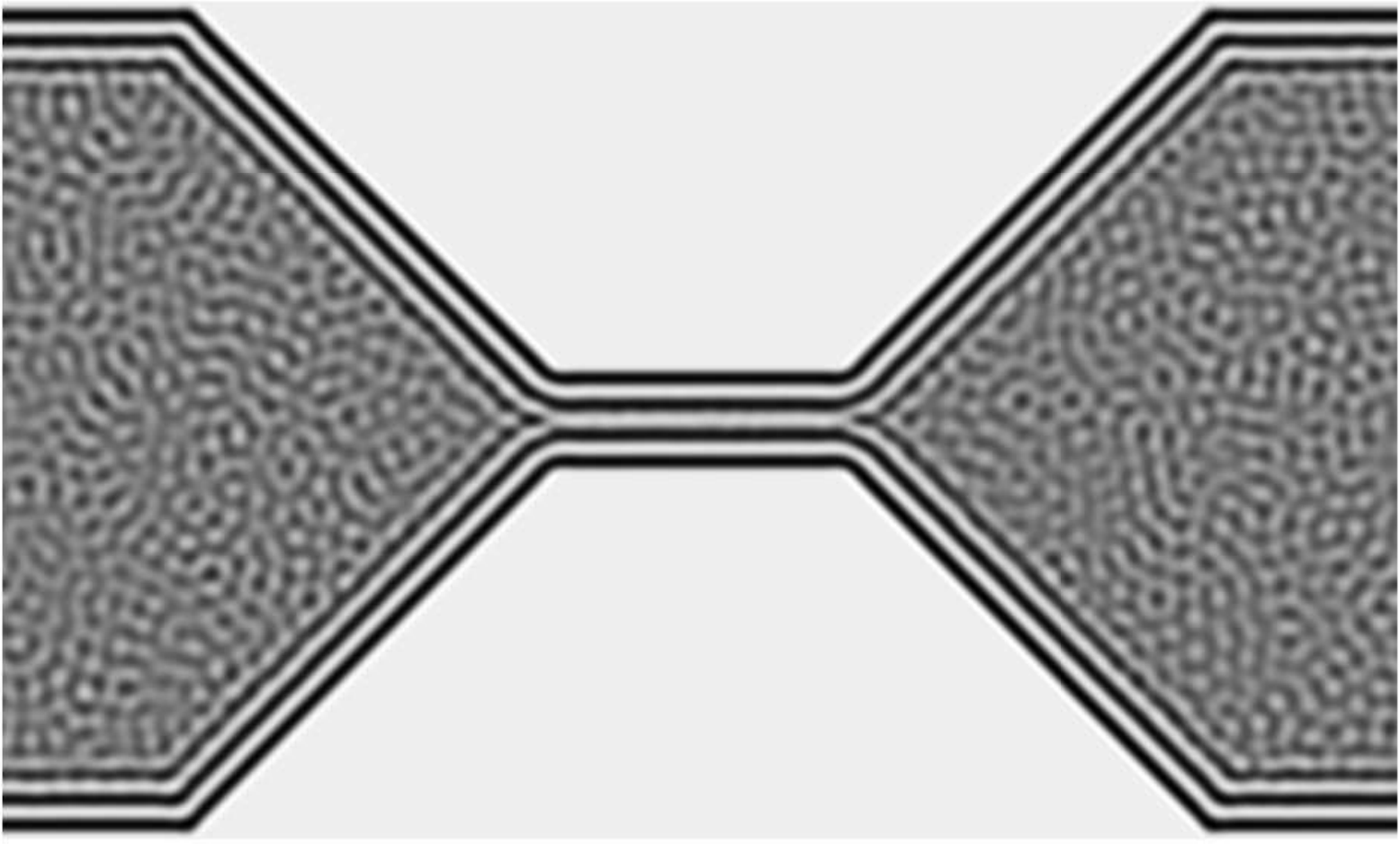}
\end{center}
\end{minipage}
\begin{minipage}{0.31\linewidth}
\begin{center}
\includegraphics[width=1.6in]{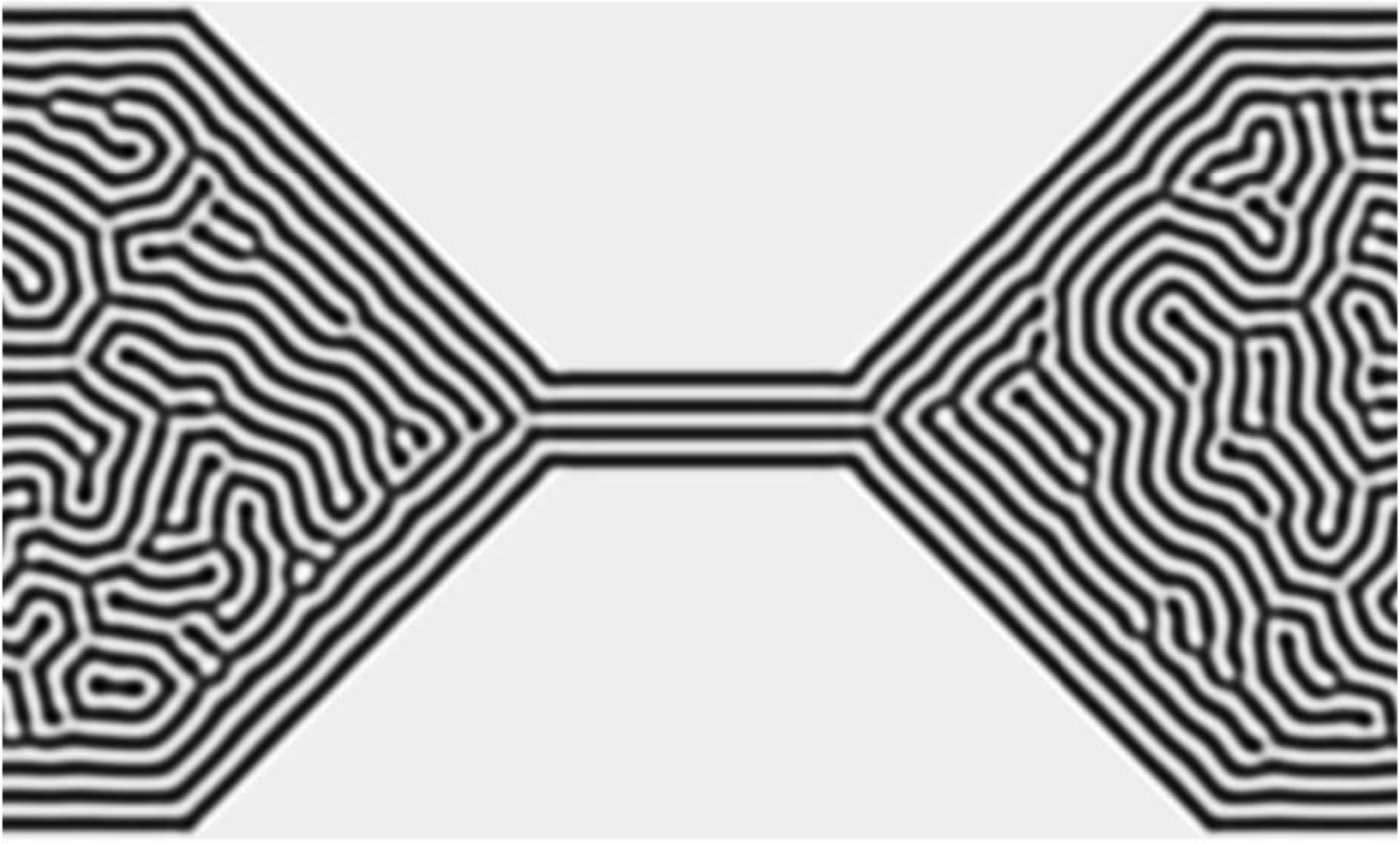}
\end{center}
\end{minipage}
\begin{minipage}{0.31\linewidth}
\begin{center}
\includegraphics[width=1.6in]{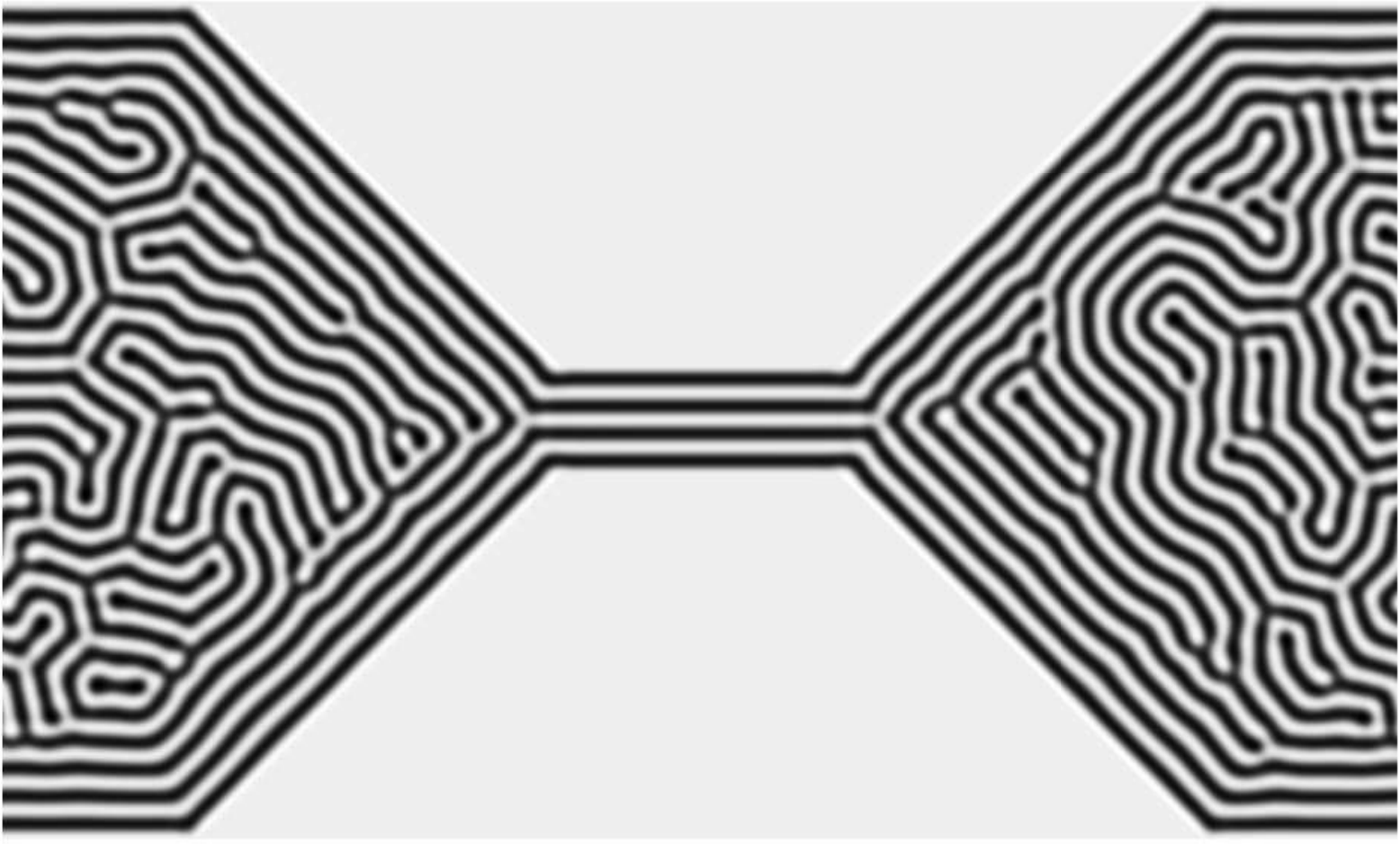}
\end{center}
\end{minipage}\\
\begin{minipage}{0.03\linewidth}
\centering (b)
\end{minipage}
\begin{minipage}{0.31\linewidth}
\begin{center}
\includegraphics[width=1.6in]{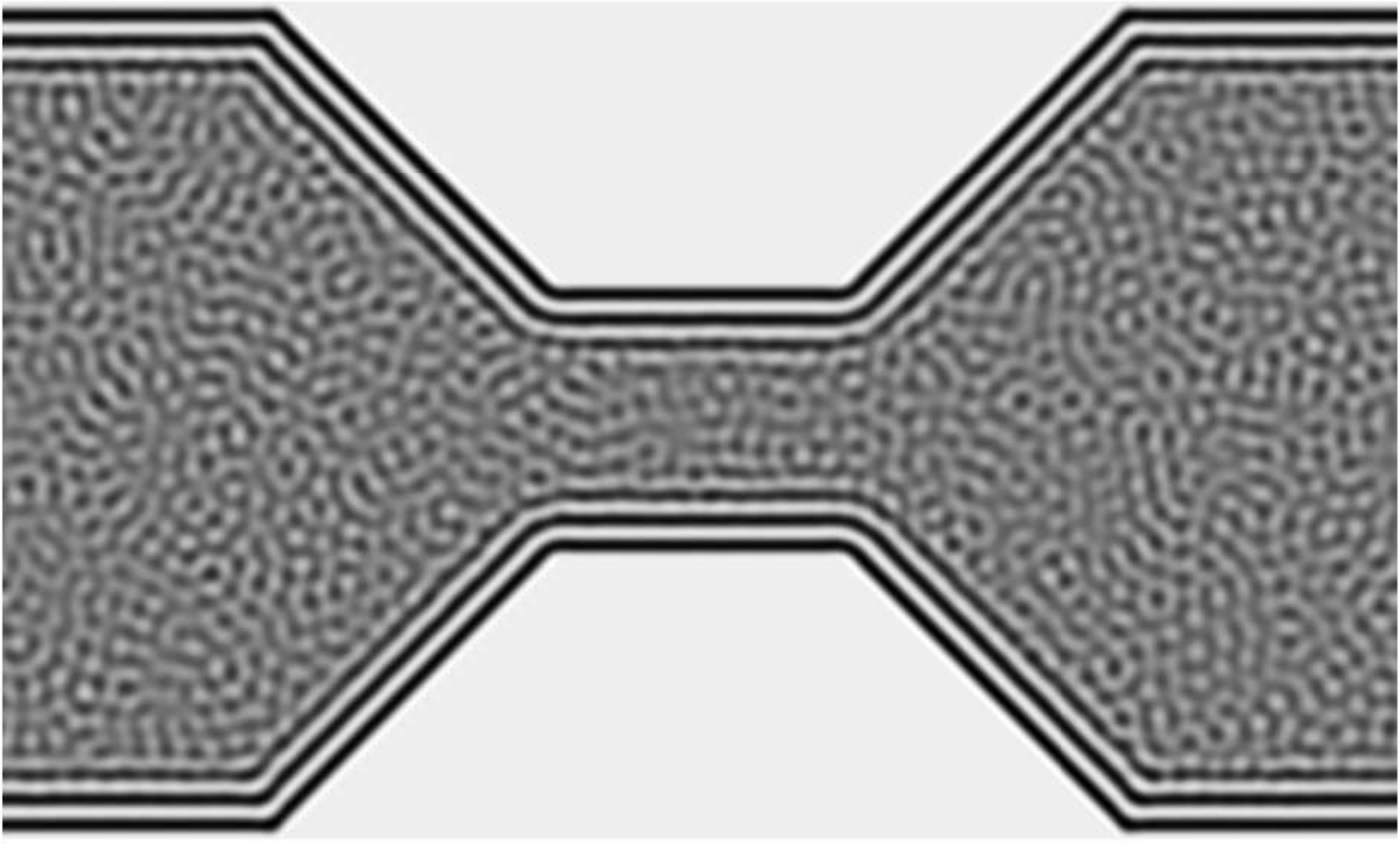}
$t = 30\Delta t$
\end{center}
\end{minipage}
\begin{minipage}{0.31\linewidth}
\begin{center}
\includegraphics[width=1.6in]{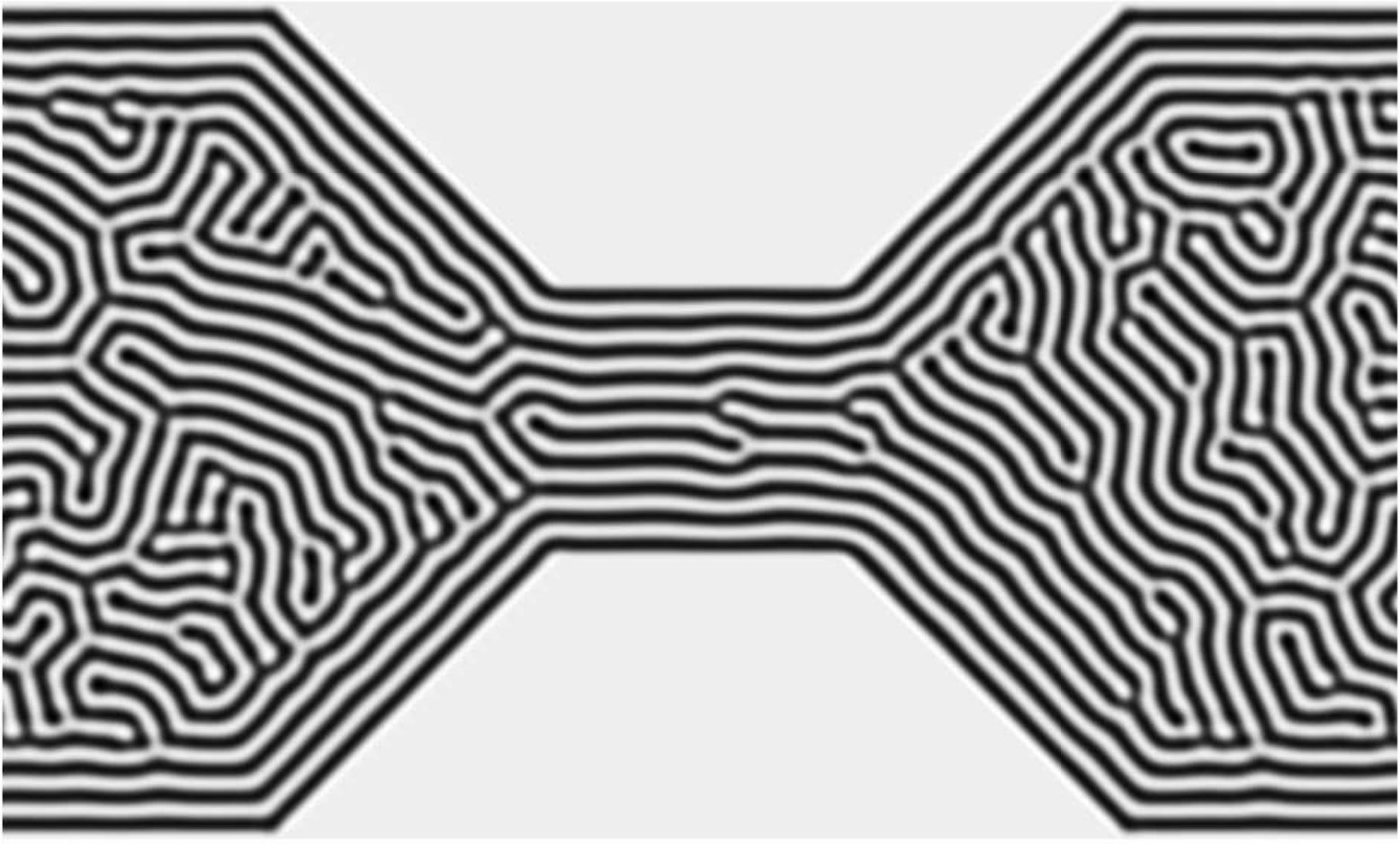}
$t = 100\Delta t$
\end{center}
\end{minipage}
\begin{minipage}{0.31\linewidth}
\begin{center}
\includegraphics[width=1.6in]{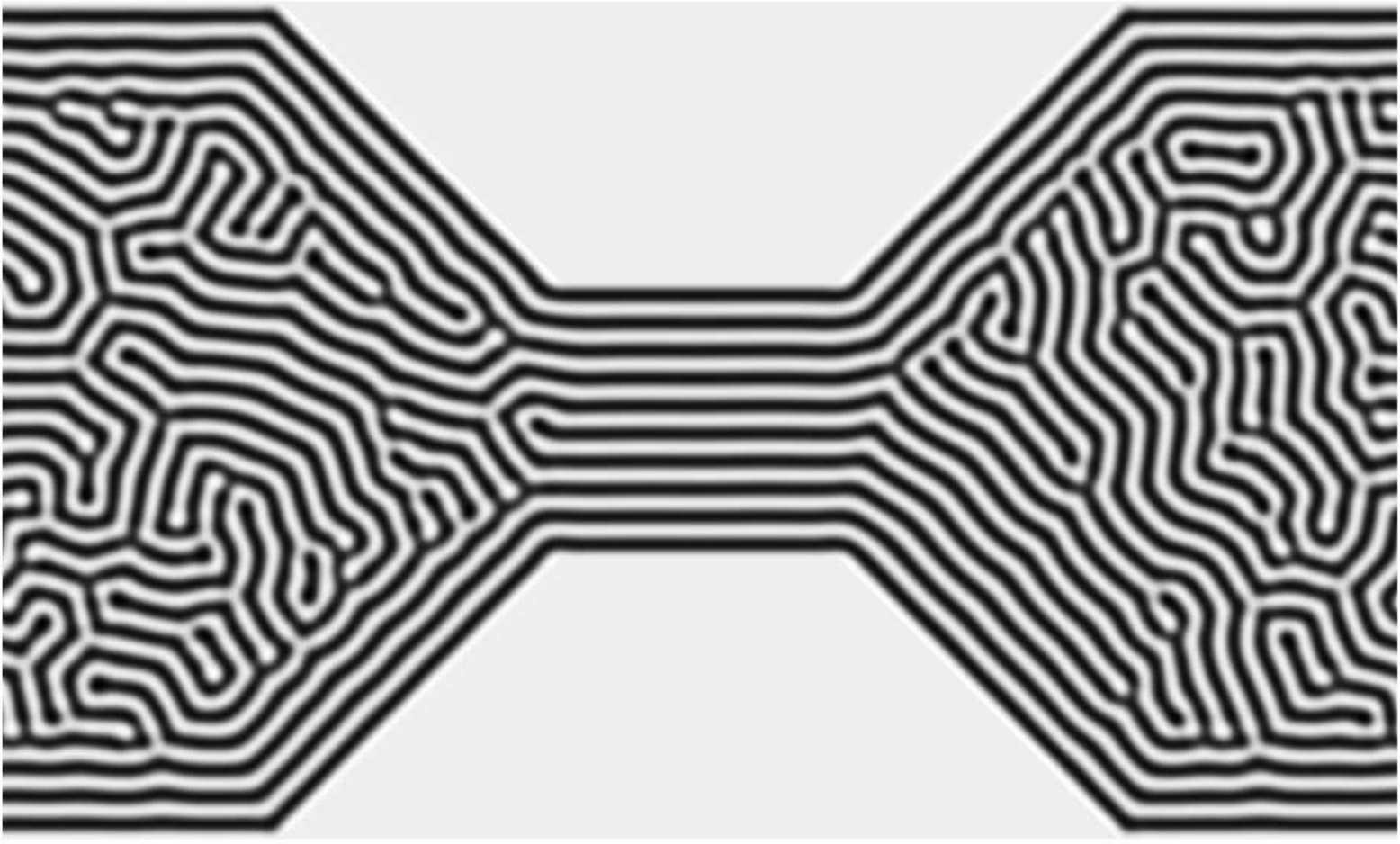}
$t = 2000\Delta t$
\end{center}
\end{minipage}
\vspace{2mm}
\caption{The effect of different trench width: (a) $4L^*$ and (b)
$10L^*$. Evolution times are given below each figure. }
\label{width}
\end{figure}

 To investigate the effect of the channel width, we fix
$a=5L^*$ with $b=2L^*$ and $b=5L^*$. Figures~\ref{width}~(a) and
(b) show the temporal evolution of $\phi$ with the trench widths
$2b=4L^*$ and $2b=10L^*$, respectively. We can observe that the
self-assembled pattern is completely defect-free and is aligned
parallel to the trench walls within the narrow trench area; all the
defects reside in the wider regions on either side, which is
consistent with the experimental results \cite{RRZS}.

\subsection{The effect of channel length}
\vspace{2mm}

\begin{figure}[!h]
\vspace{2mm}
\begin{minipage}{0.03\linewidth}
\centering (a)
\end{minipage}
\begin{minipage}{0.31\linewidth}
\begin{center}
\includegraphics[width=1.6in]{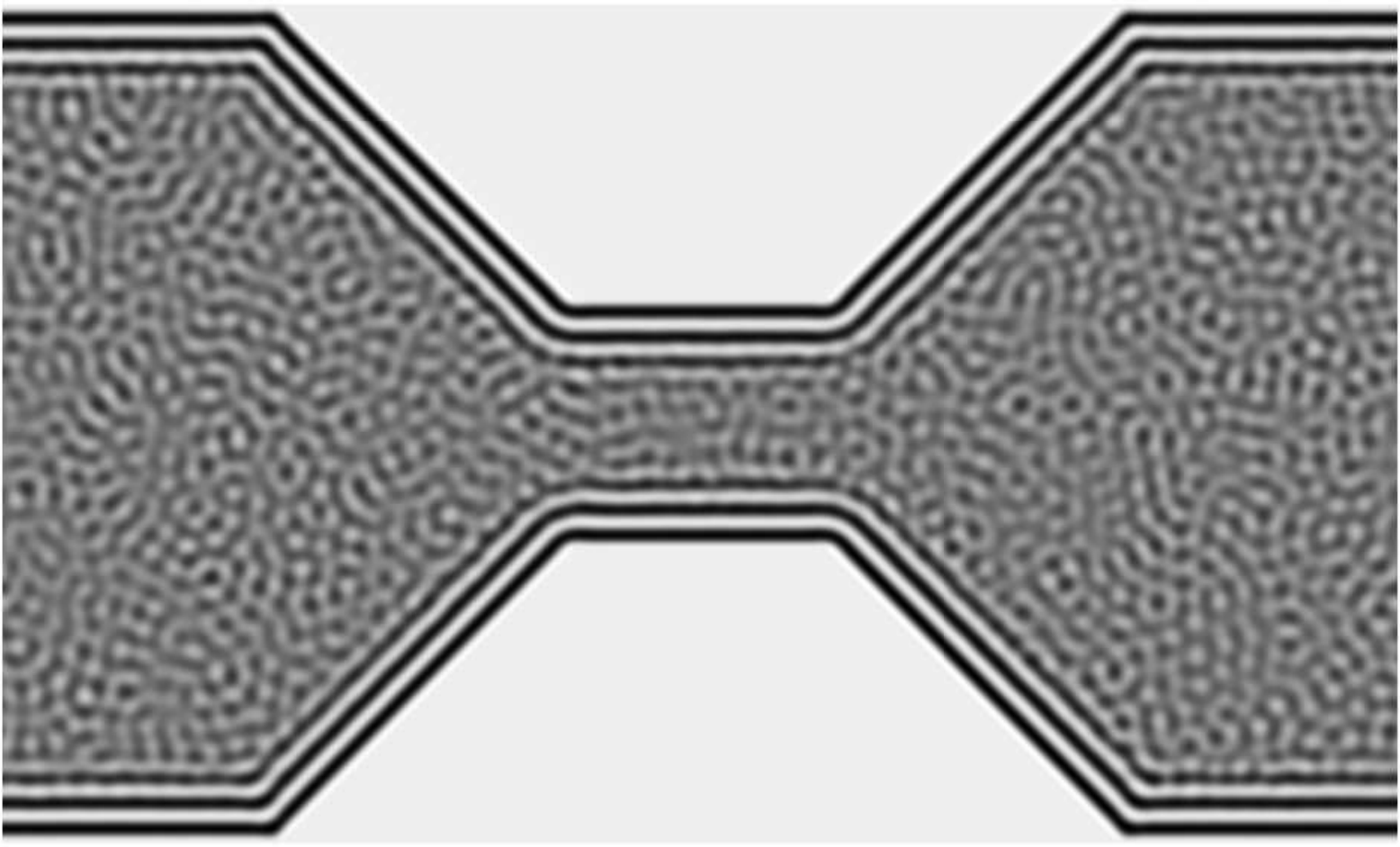}
\end{center}
\end{minipage}
\begin{minipage}{0.31\linewidth}
\begin{center}
\includegraphics[width=1.6in]{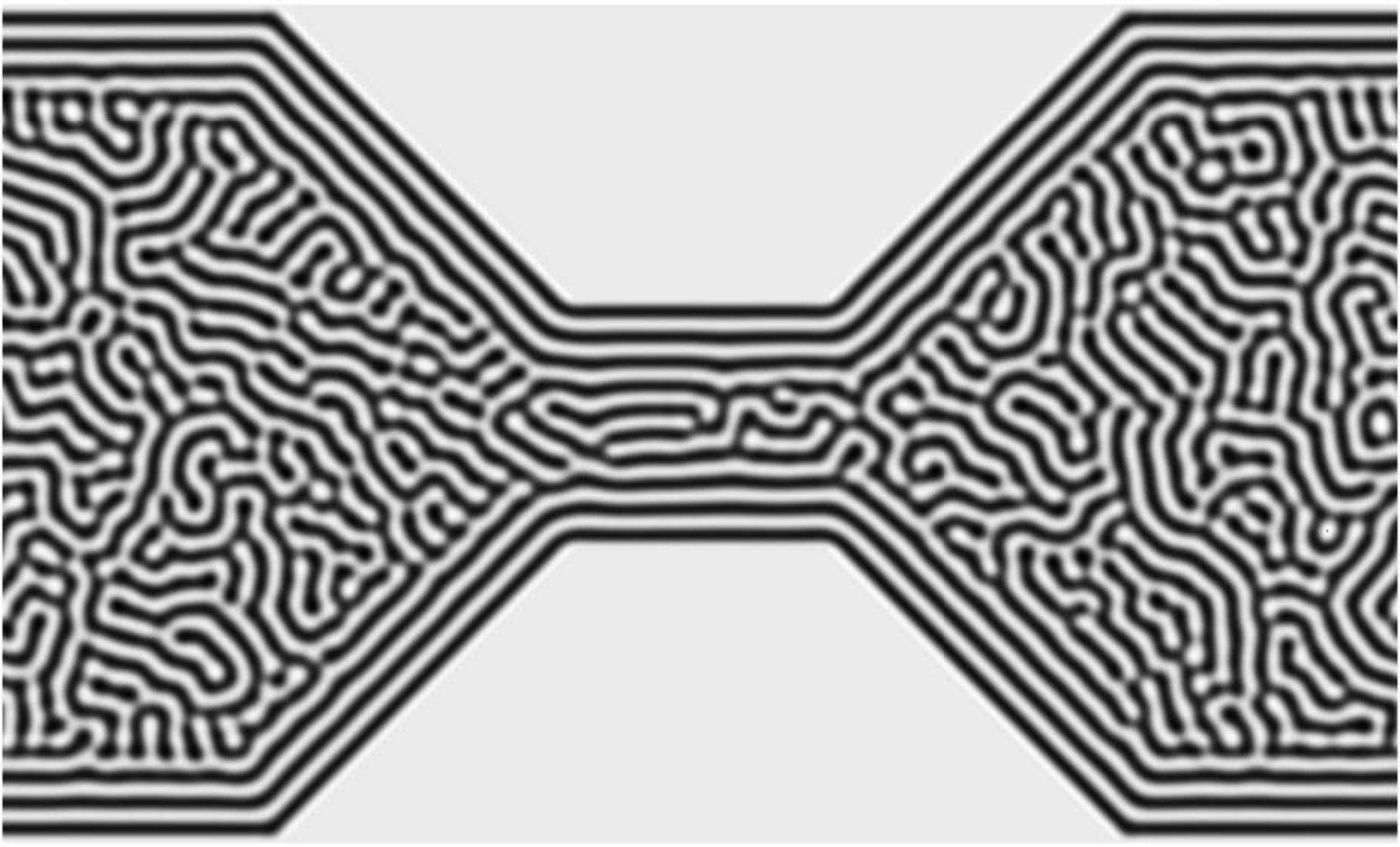}
\end{center}
\end{minipage}
\begin{minipage}{0.31\linewidth}
\begin{center}
\includegraphics[width=1.6in]{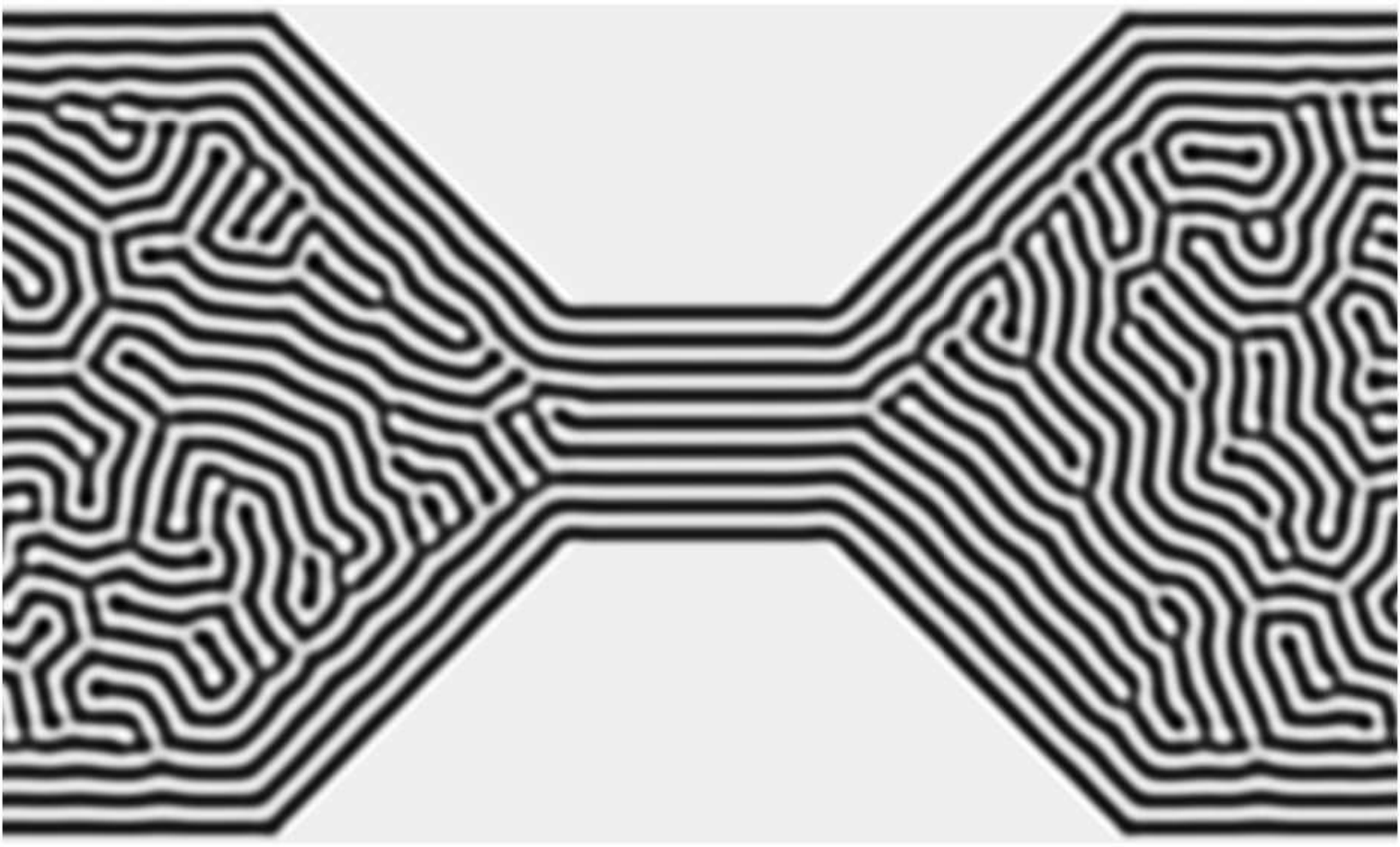}
\end{center}
\end{minipage}\\
\begin{minipage}{0.03\linewidth}
\centering (b)
\end{minipage}
\begin{minipage}{0.31\linewidth}
\begin{center}
\includegraphics[width=1.6in]{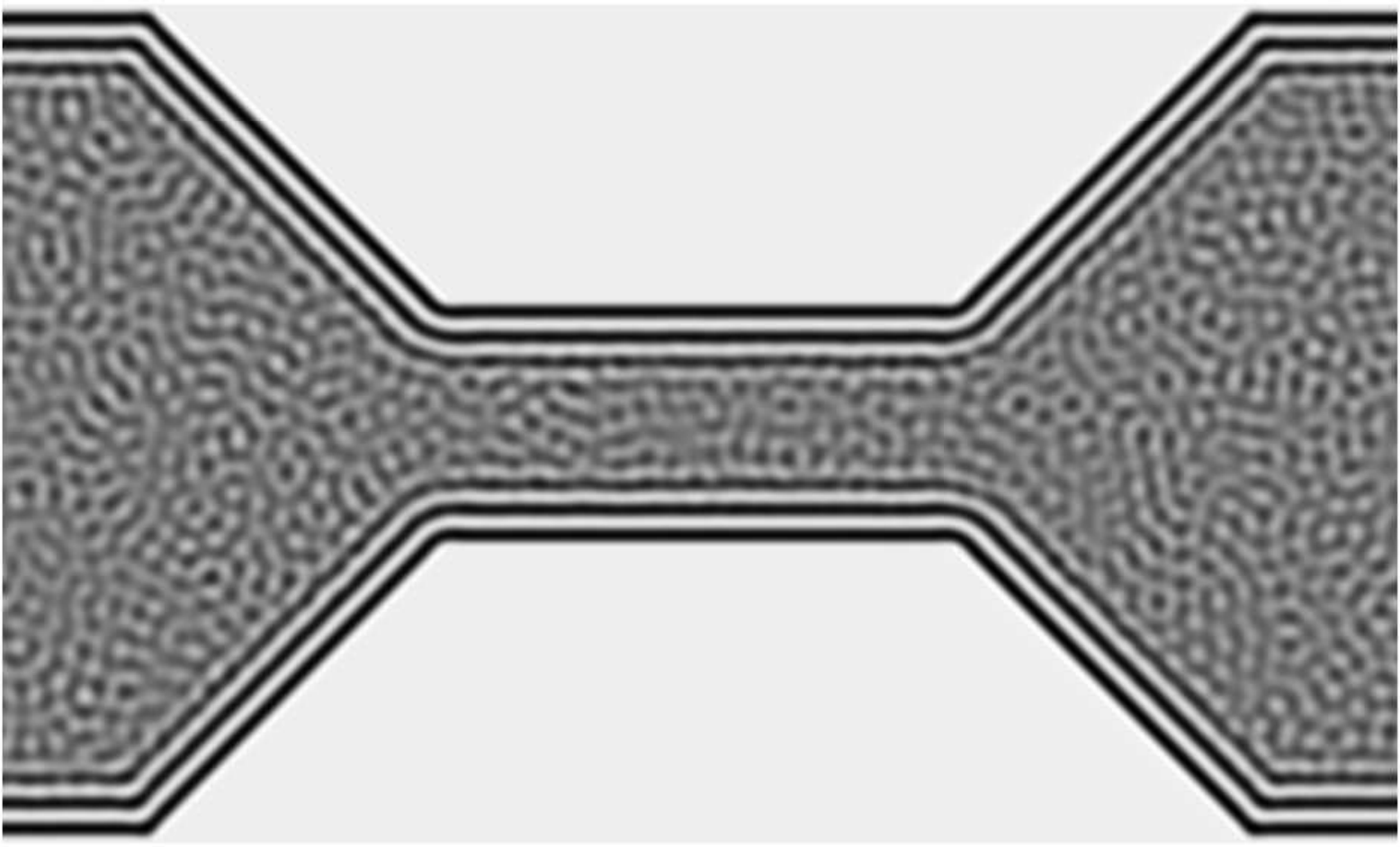} \\
$t = 30\Delta t$
\end{center}
\end{minipage}
\begin{minipage}{0.31\linewidth}
\begin{center}
\includegraphics[width=1.6in]{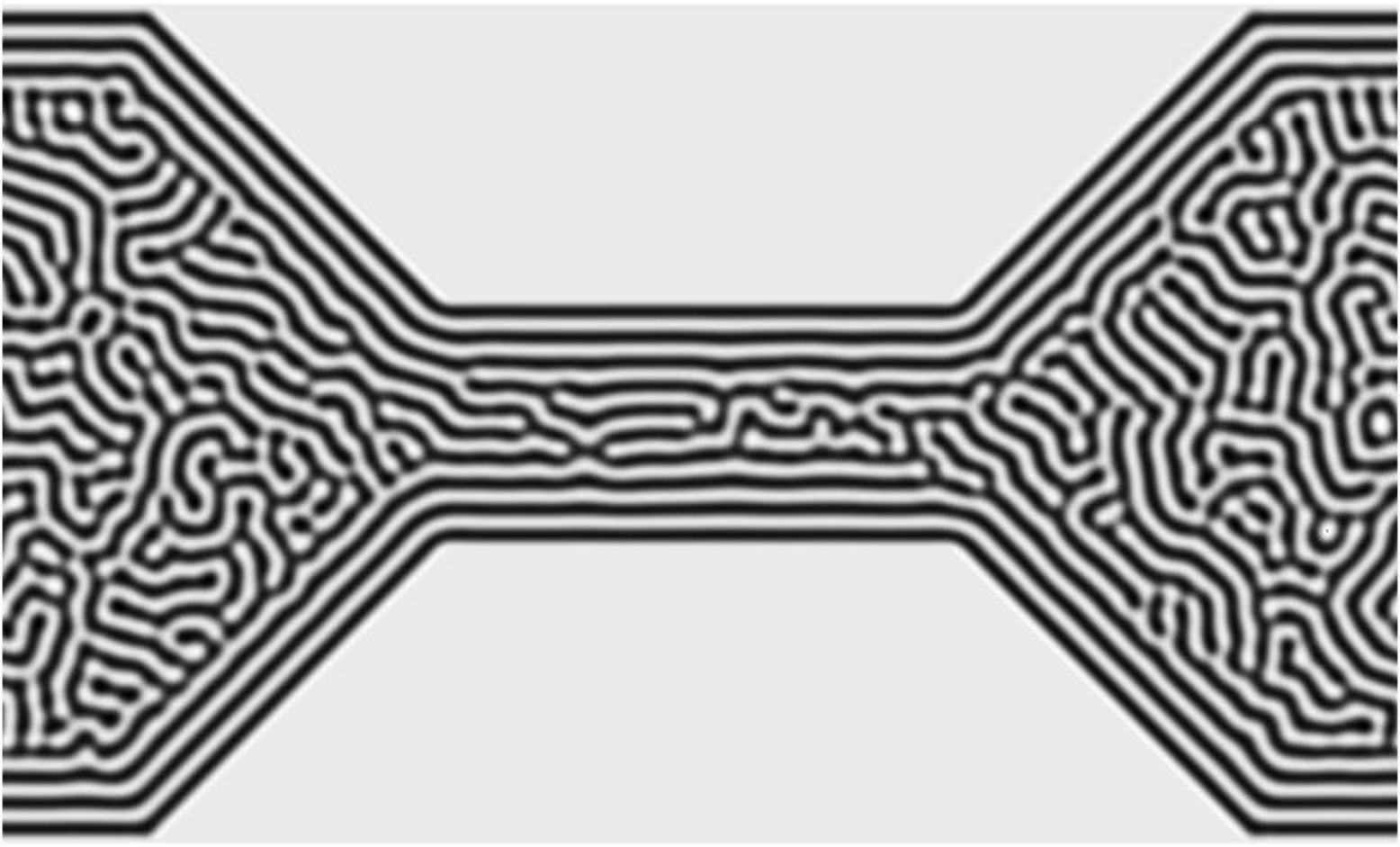} \\
$t = 100\Delta t$
\end{center}
\end{minipage}
\begin{minipage}{0.31\linewidth}
\begin{center}
\includegraphics[width=1.6in]{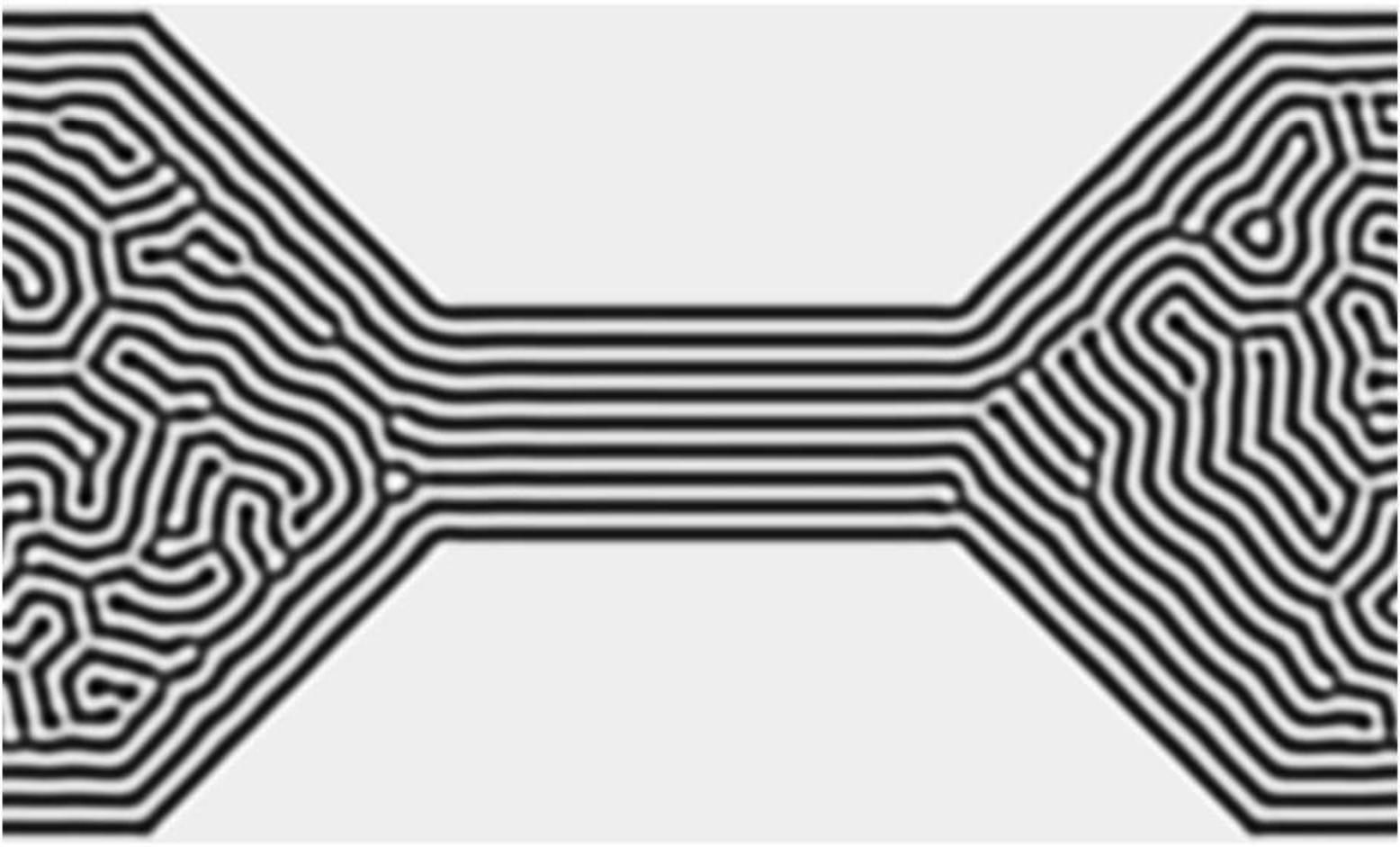} \\
$t = 2000\Delta t$
\end{center}
\end{minipage}
\vspace{2mm}
\caption{The effect of different trench length: (a) $9L^*$ and (b)
$18L^*$. Evolution times are given below each figure. }
\label{lengths}
\end{figure}

 In this section, we simulate two cases with respect to a
narrow channel length. For this test, we use two different values
$a=4.5L^*$ and $a=9L^*$ when we fix $b=4.5L^*$. The numerical
results can be seen in figure~\ref{lengths}. Similarly to the
previous tests, we can see that the numerical solution in the narrow
channel has the defect-free lamella pattern.

\subsection{The effect of angle}
\vspace{2mm}

\begin{figure}[!t]
\vspace{2mm}
\begin{minipage}{0.03\linewidth}
\centering (a)
\end{minipage}
\begin{minipage}{0.31\linewidth}
\begin{center}
\includegraphics[clip=true,width=1.6in]{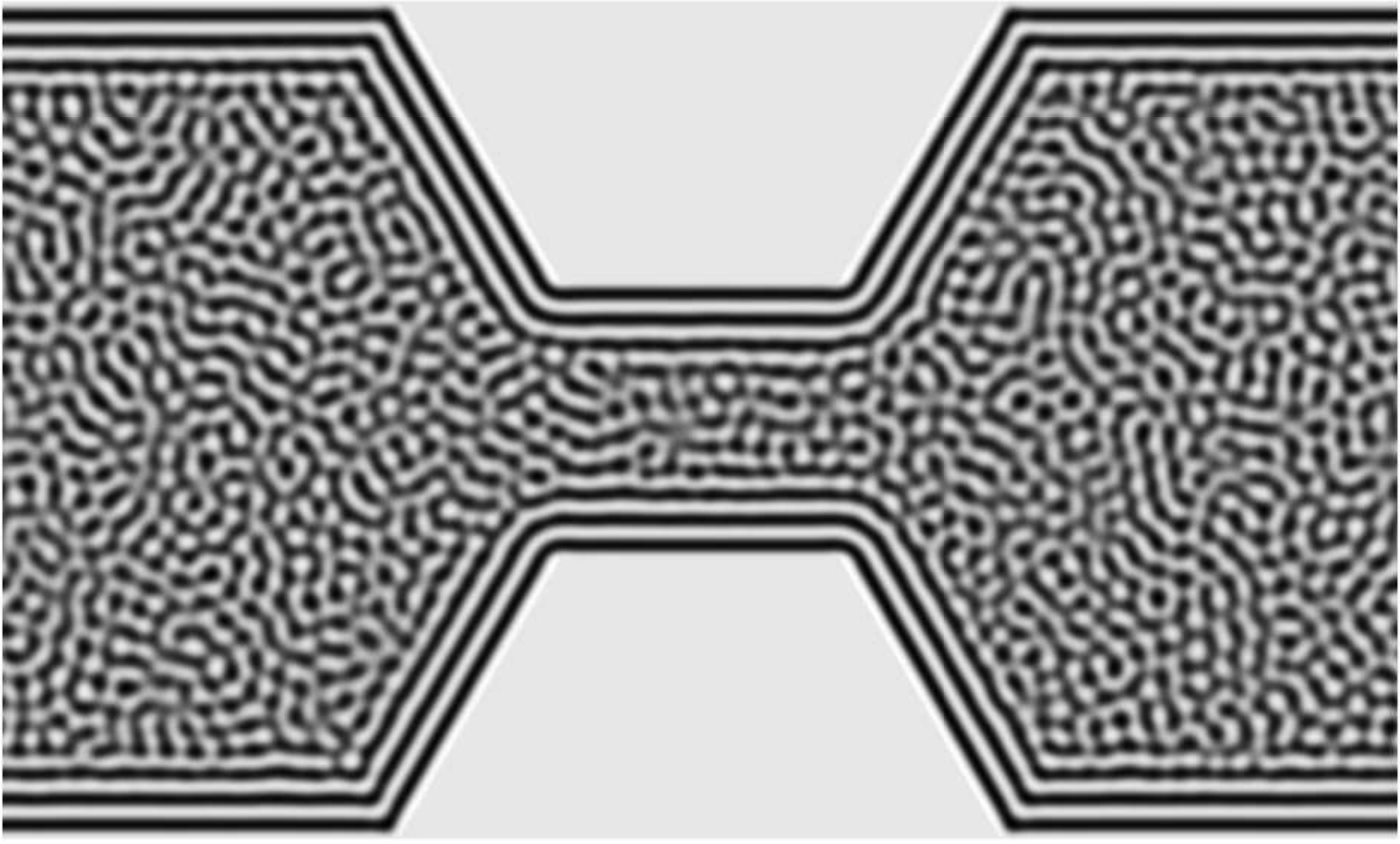}
\end{center}
\end{minipage}
\begin{minipage}{0.31\linewidth}
\begin{center}
\includegraphics[clip=true,width=1.6in]{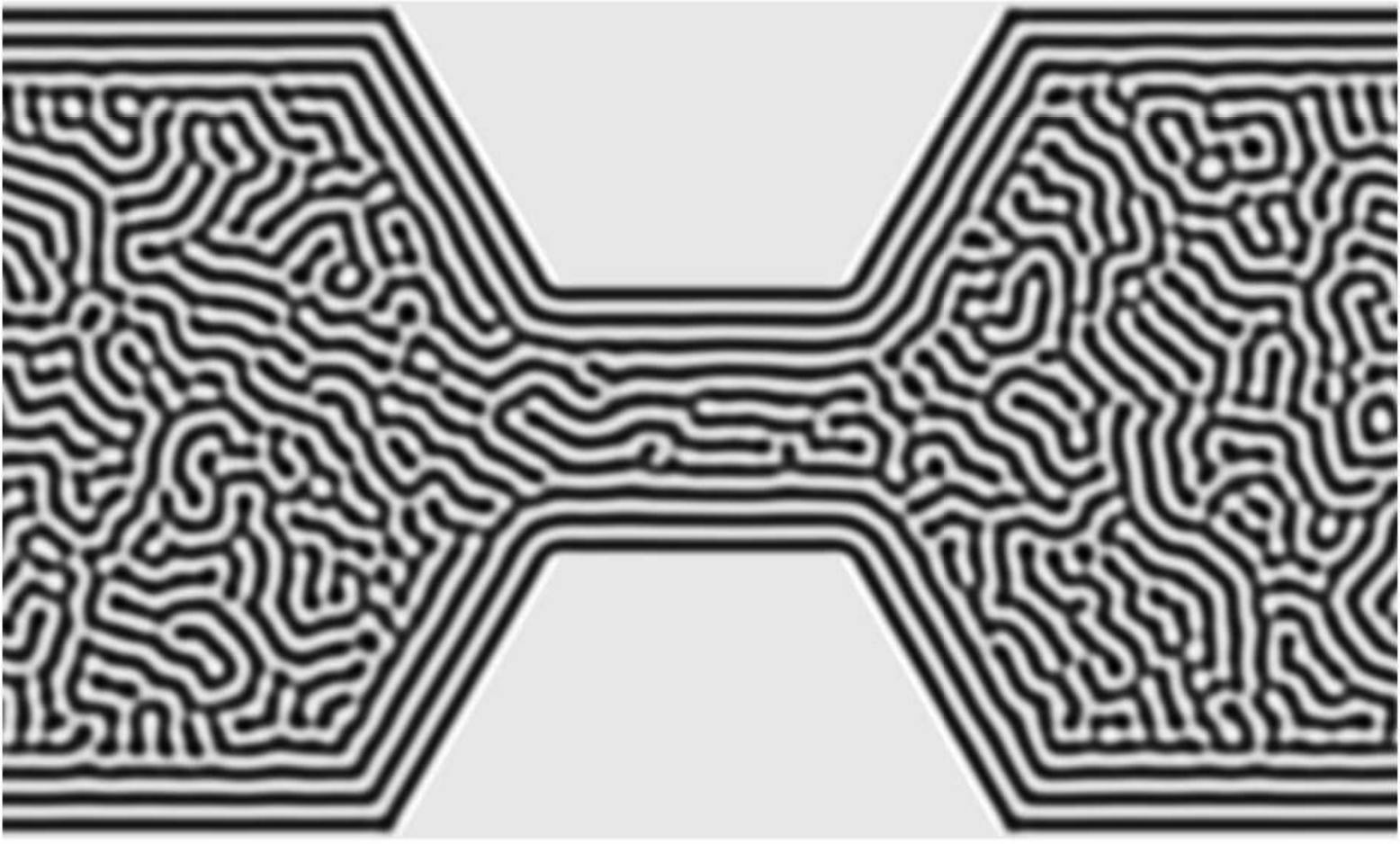}
\end{center}
\end{minipage}
\begin{minipage}{0.31\linewidth}
\begin{center}
\includegraphics[clip=true,width=1.6in]{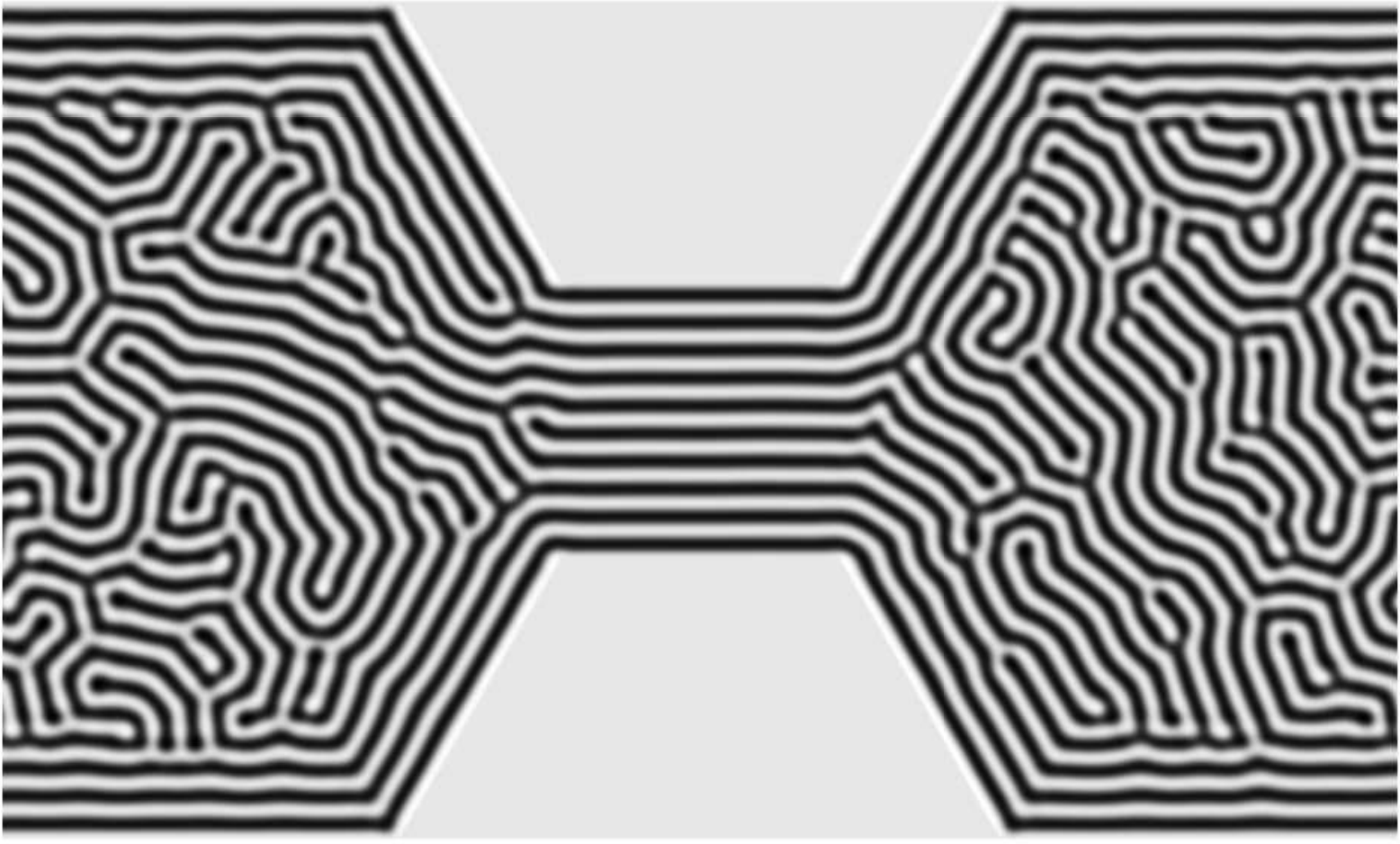}
\end{center}
\end{minipage}\\
\begin{minipage}{0.03\linewidth}
\centering (b)
\end{minipage}
\begin{minipage}{0.31\linewidth}
\begin{center}
\includegraphics[clip=true,width=1.6in]{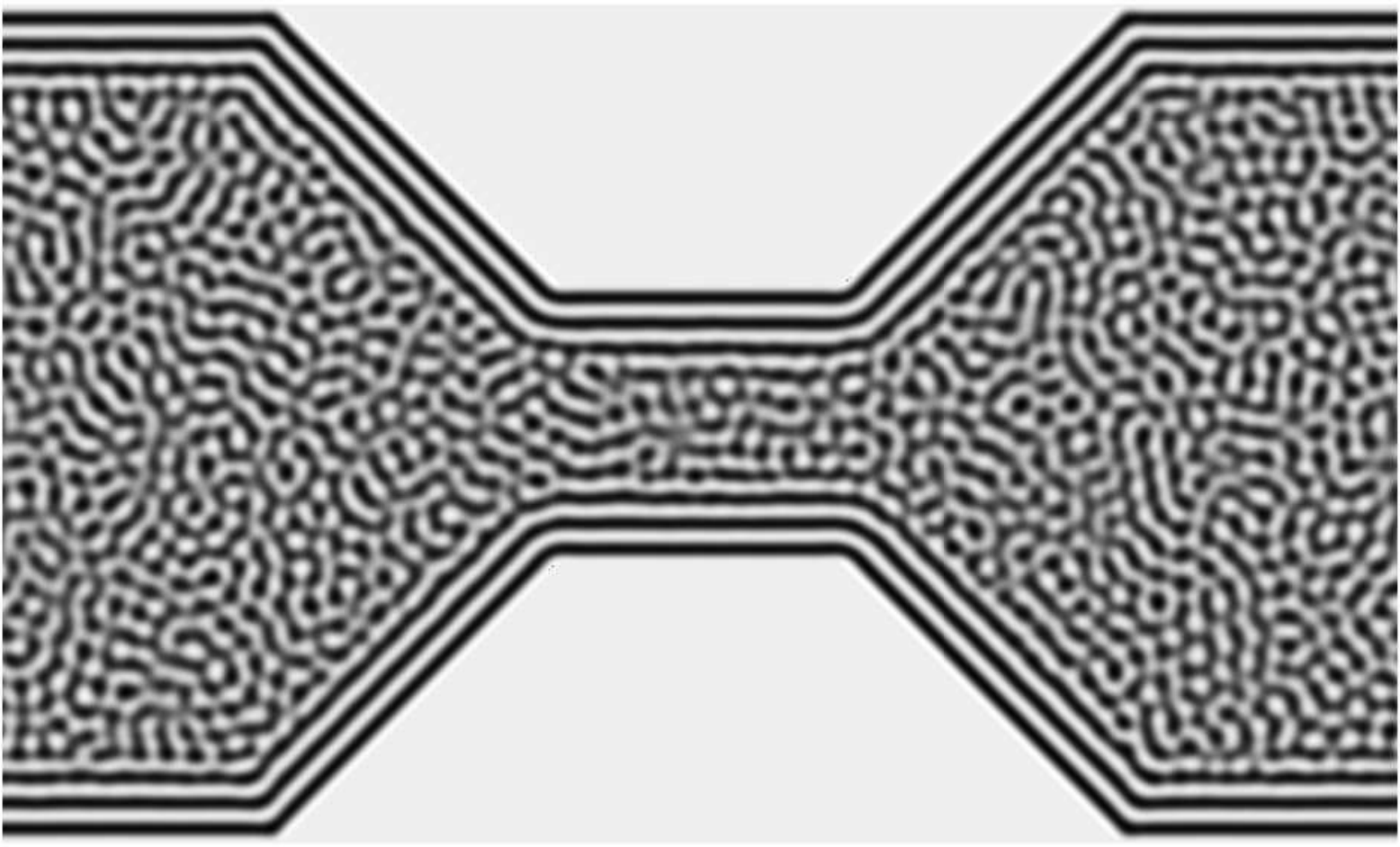}
\end{center}
\end{minipage}
\begin{minipage}{0.31\linewidth}
\begin{center}
\includegraphics[clip=true,width=1.6in]{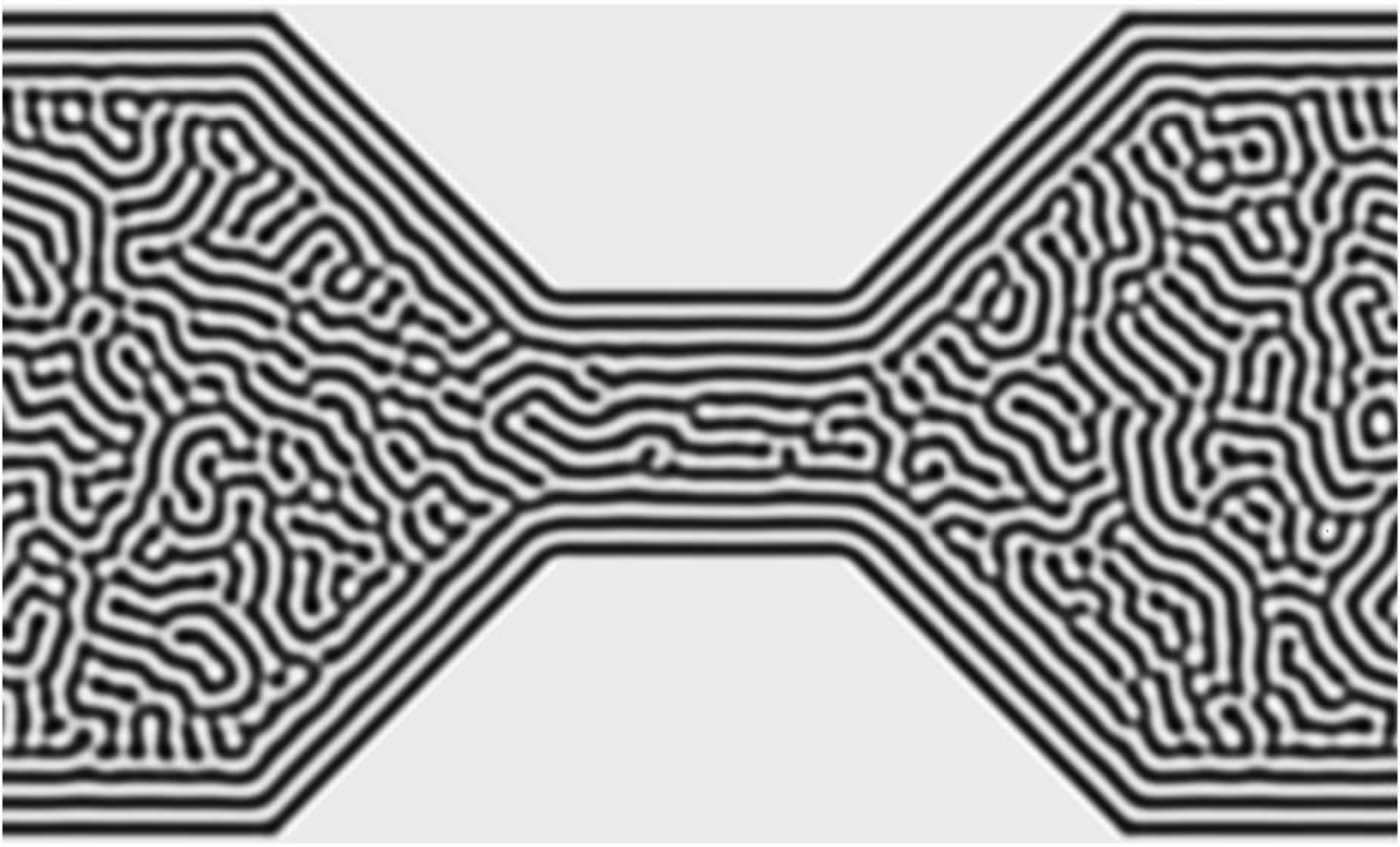}
\end{center}
\end{minipage}
\begin{minipage}{0.31\linewidth}
\begin{center}
\includegraphics[clip=true,width=1.6in]{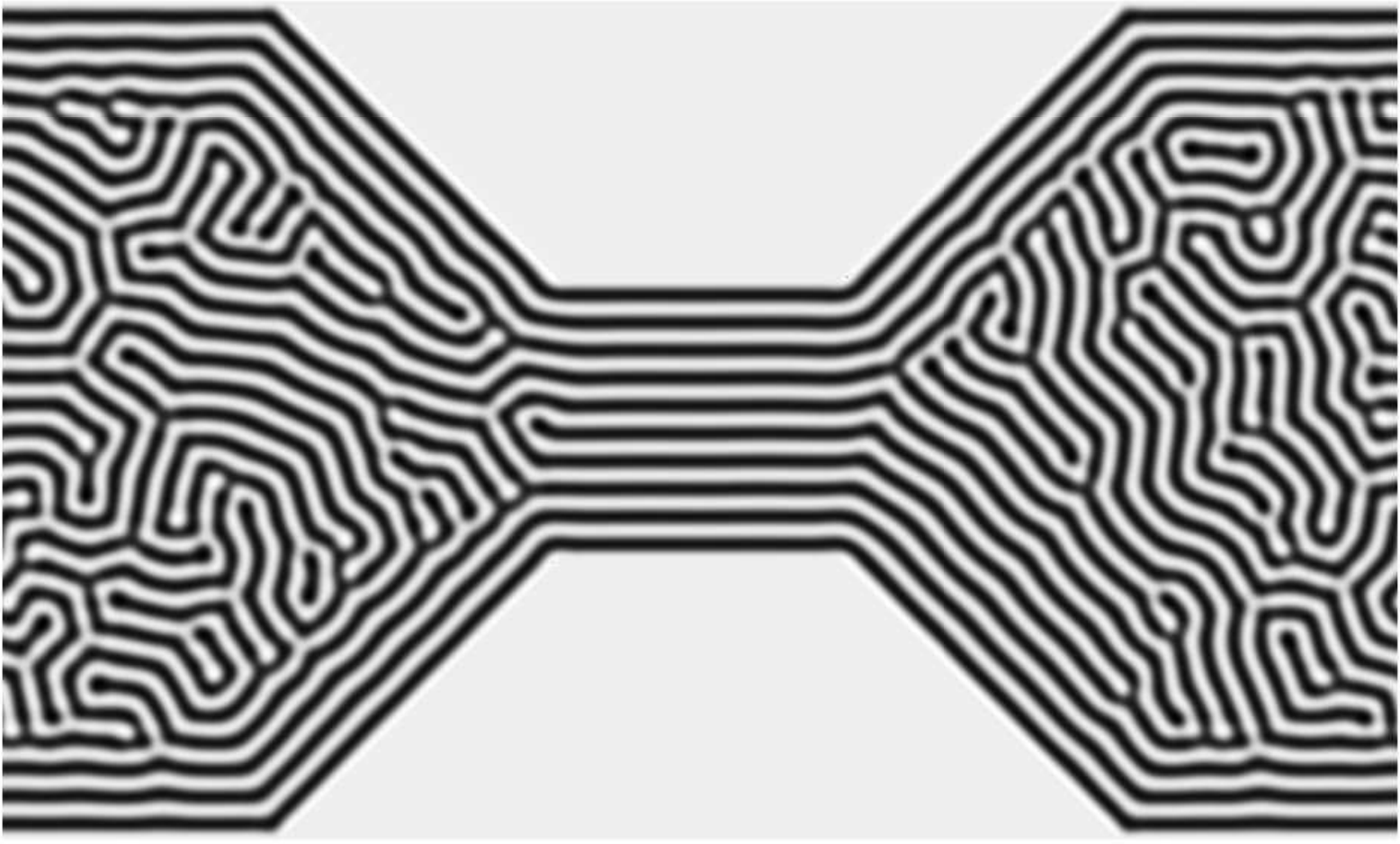}
\end{center}
\end{minipage}\\
\begin{minipage}{0.03\linewidth}
\centering (c)
\end{minipage}
\begin{minipage}{0.31\linewidth}
\begin{center}
\includegraphics[clip=true,width=1.6in]{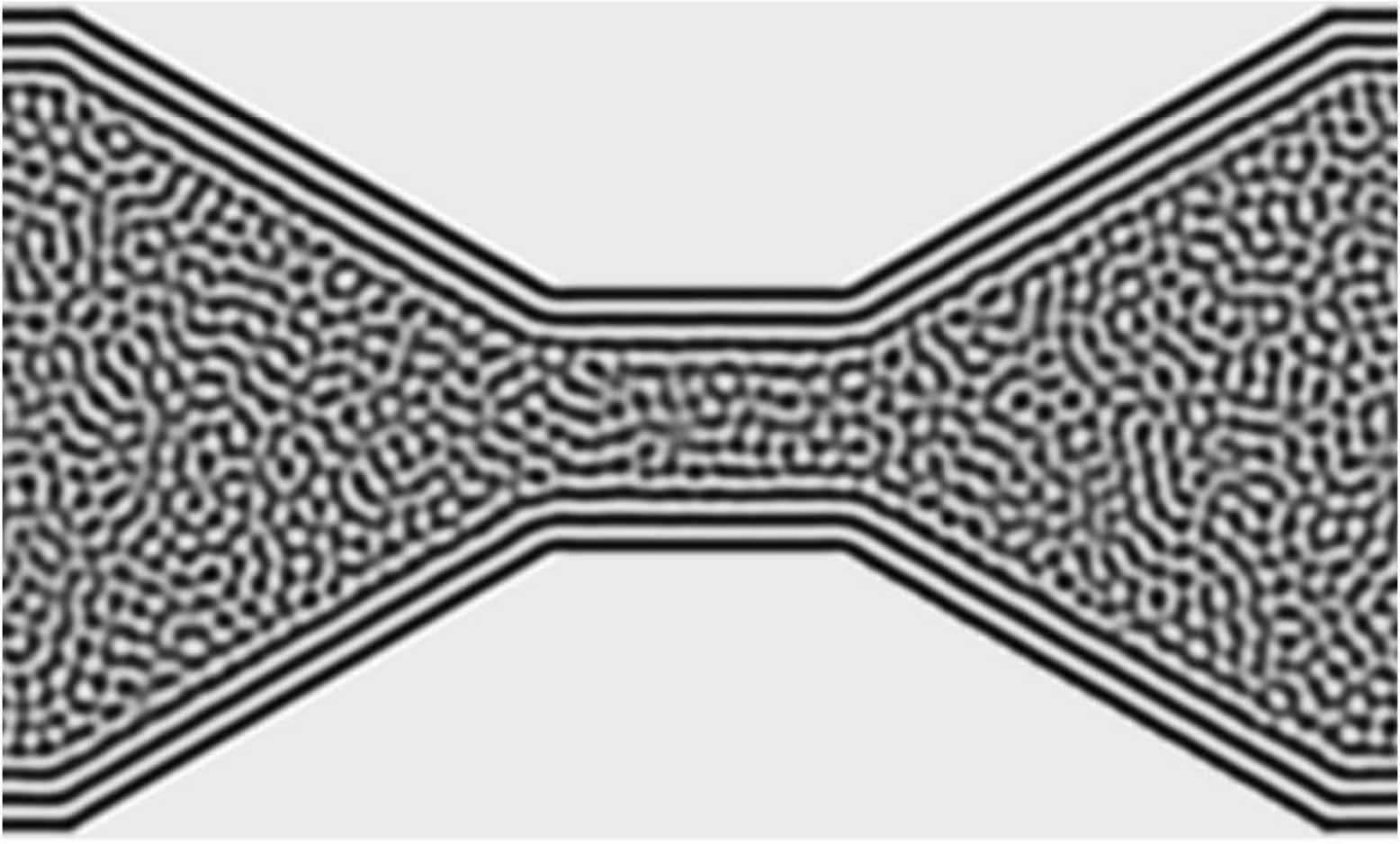}\\
$t = 40\Delta t$
\end{center}
\end{minipage}
\begin{minipage}{0.31\linewidth}
\begin{center}
\includegraphics[clip=true,width=1.6in]{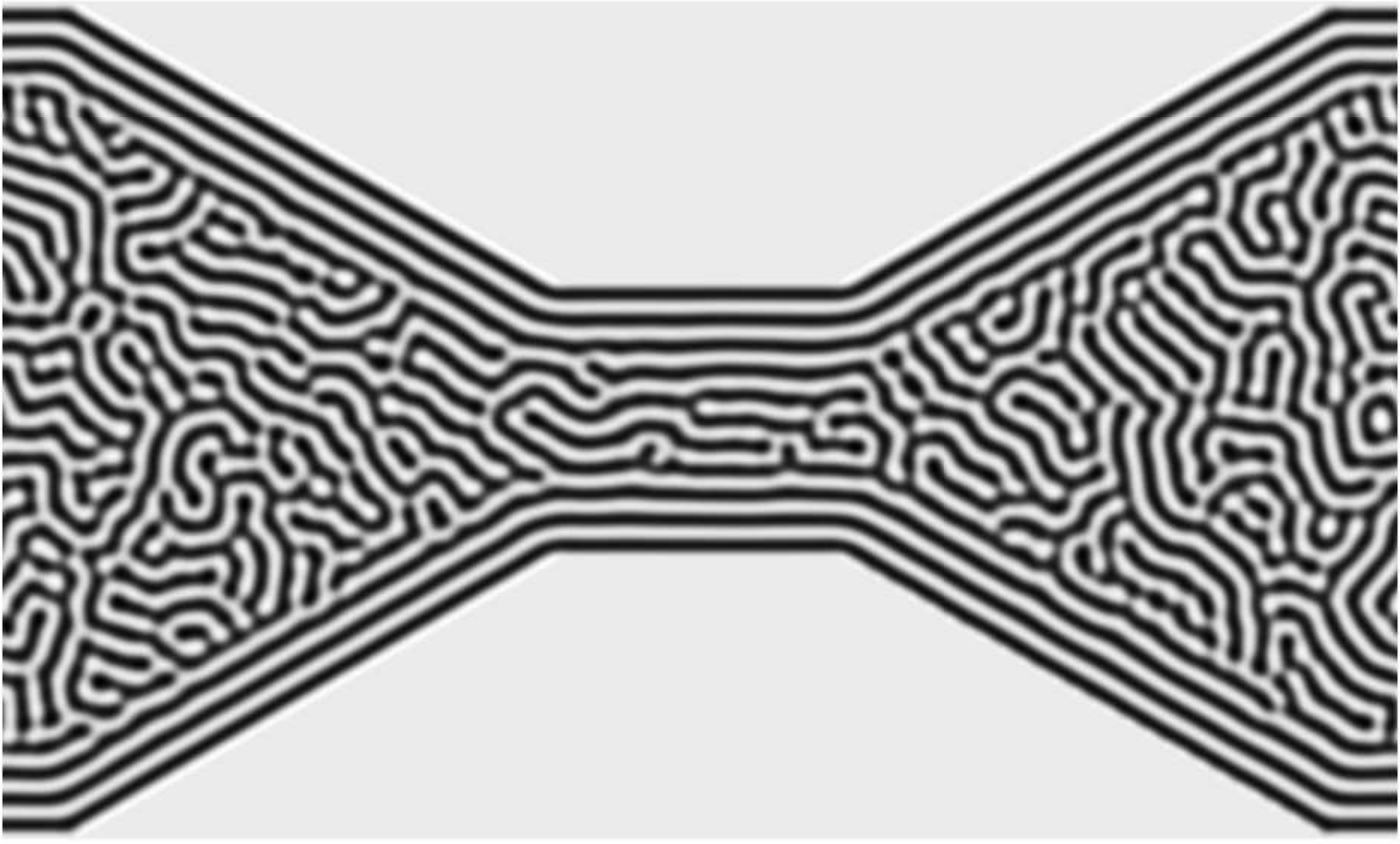}\\
$t = 100\Delta t$
\end{center}
\end{minipage}
\begin{minipage}{0.31\linewidth}
\begin{center}
\includegraphics[clip=true,width=1.6in]{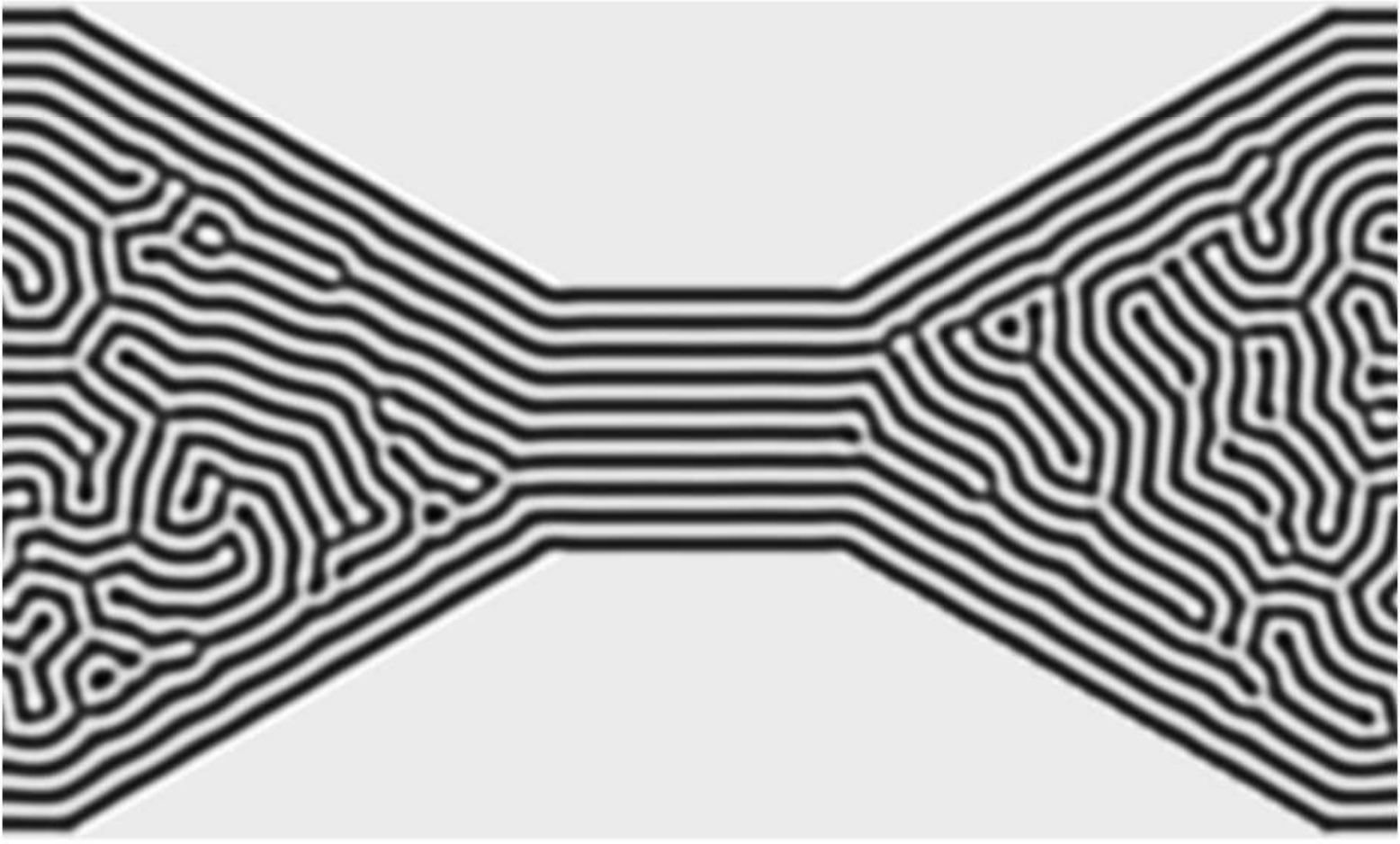}\\
$t = 2000\Delta t$
\end{center}
\end{minipage}
\vspace{2mm}
\caption{The effect of the angle: (a) $\theta = \pi/3$, (b) $\pi/4$, and (c)
$\pi/6$. Evolution times are given below each figure.}
\label{angle}
\end{figure}

To see the dynamics of the angle, we only change the angle as $\theta =
\pi/3$, $\pi/4$, and $\pi/6$ with $a=b=5L^*$. Figure \ref{angle}
represents the temporal evolution of pattern formation in channels
with respect to the angle. In all three cases, we observe that the
numerical solution in the narrow channel has aligned lamella
patterns parallel to the trench walls. Also, within the narrow
trench region, the self-assembled pattern is defect-free unlike
the side region where all the defects are located.


\begin{figure}[!b]
\begin{minipage}{1.0\linewidth}
\begin{center}
\includegraphics[height=2.2in]{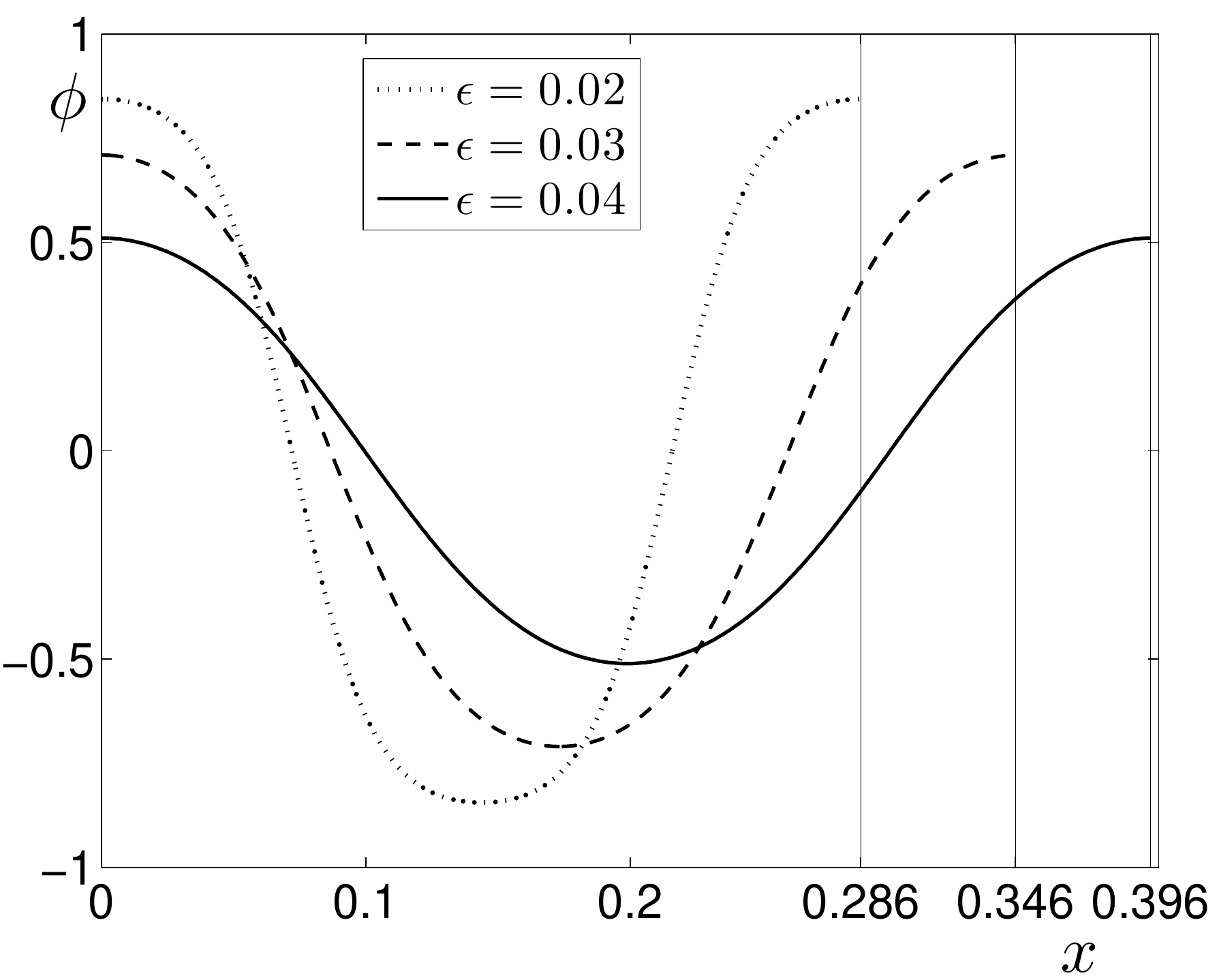}
\end{center}
\end{minipage}
\caption{Profiles of $\phi$ at the equilibrium state when $\epsilon
= 0.02$, $0.03$, and $0.04$. We reprinted from \cite{DJYY2014}, with
permission from the Current Applied Physics.} \label{Effect_EPS}
\end{figure}

Figure~\ref{Effect_EPS} shows the profiles of $\phi$ at equilibrium
state for each $\epsilon = 0.02$, $0.03$, and $0.04$.

From the result in figure~\ref{Effect_EPS}, as $\epsilon$ value is
increasing, we observe that the amplitude of $\phi$ is smaller and
the wavelength is wider.


\subsection{Comparison of Dirichlet and Neumann boundary conditions}

\begin{figure}[!h]
\begin{minipage}{0.03\linewidth}
\centering (a)
\end{minipage}
\begin{minipage}{0.31\linewidth}
\begin{center}
\includegraphics[clip=true,width=1.6in]{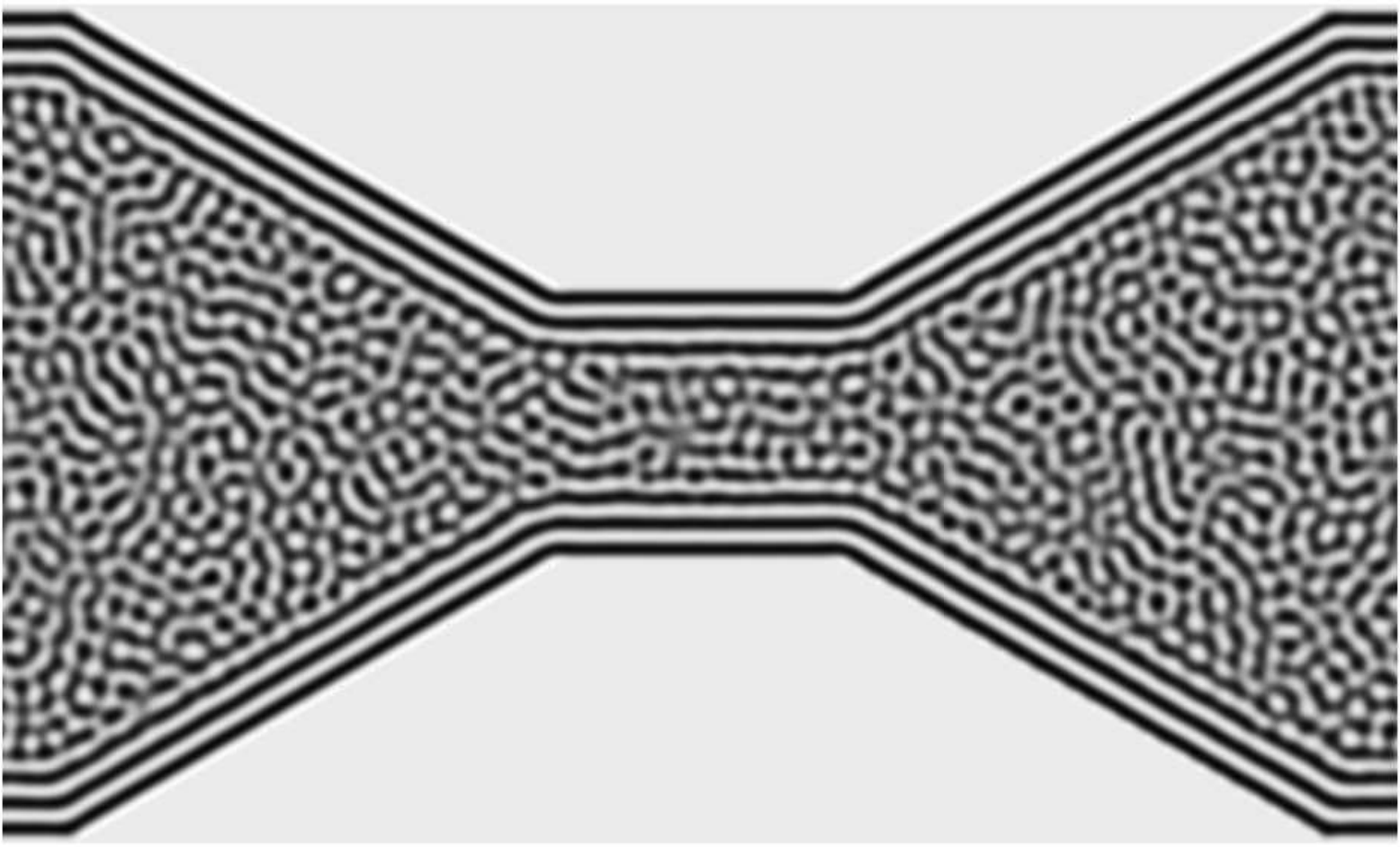}
\end{center}
\end{minipage}
\begin{minipage}{0.31\linewidth}
\begin{center}
\includegraphics[clip=true,width=1.6in]{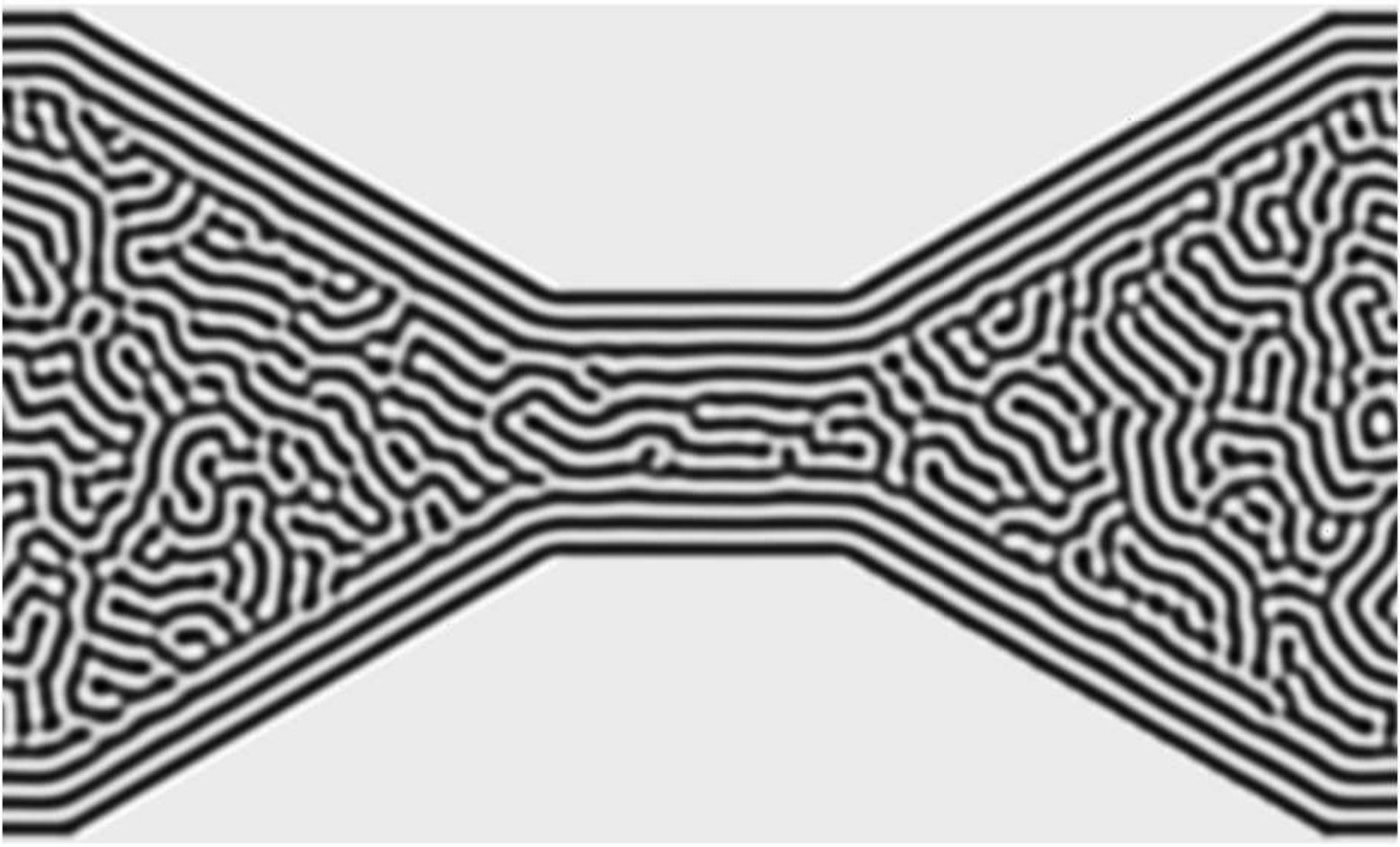}
\end{center}
\end{minipage}
\begin{minipage}{0.31\linewidth}
\begin{center}
\includegraphics[clip=true,width=1.6in]{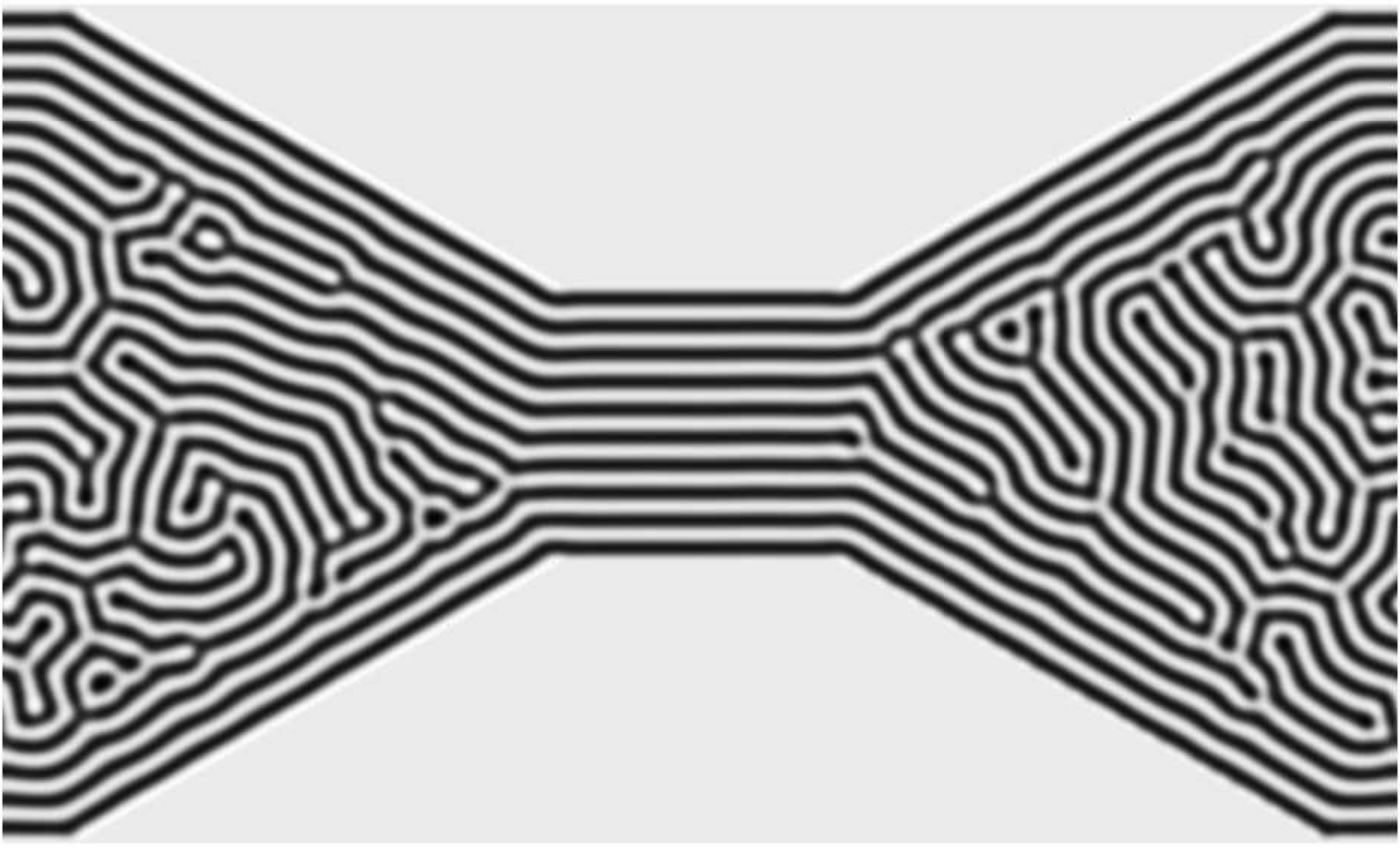}
\end{center}
\end{minipage}\\
\begin{minipage}{0.03\linewidth}
\centering (b)
\end{minipage}
\begin{minipage}{0.31\linewidth}
\begin{center}
\includegraphics[clip=true,width=1.6in]{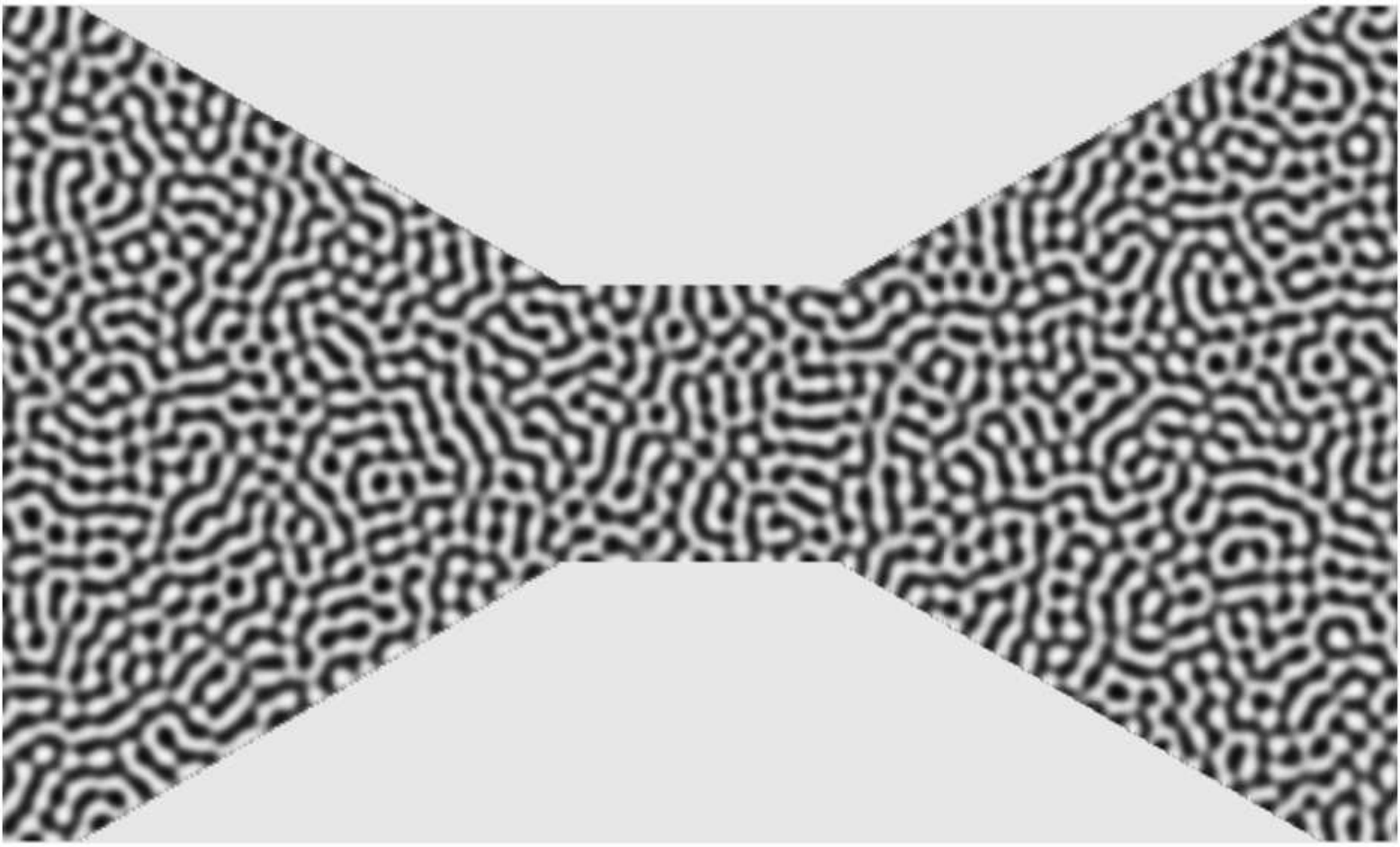}\\
$t=40 \Delta t$
\end{center}
\end{minipage}
\begin{minipage}{0.31\linewidth}
\begin{center}
\includegraphics[clip=true,width=1.6in]{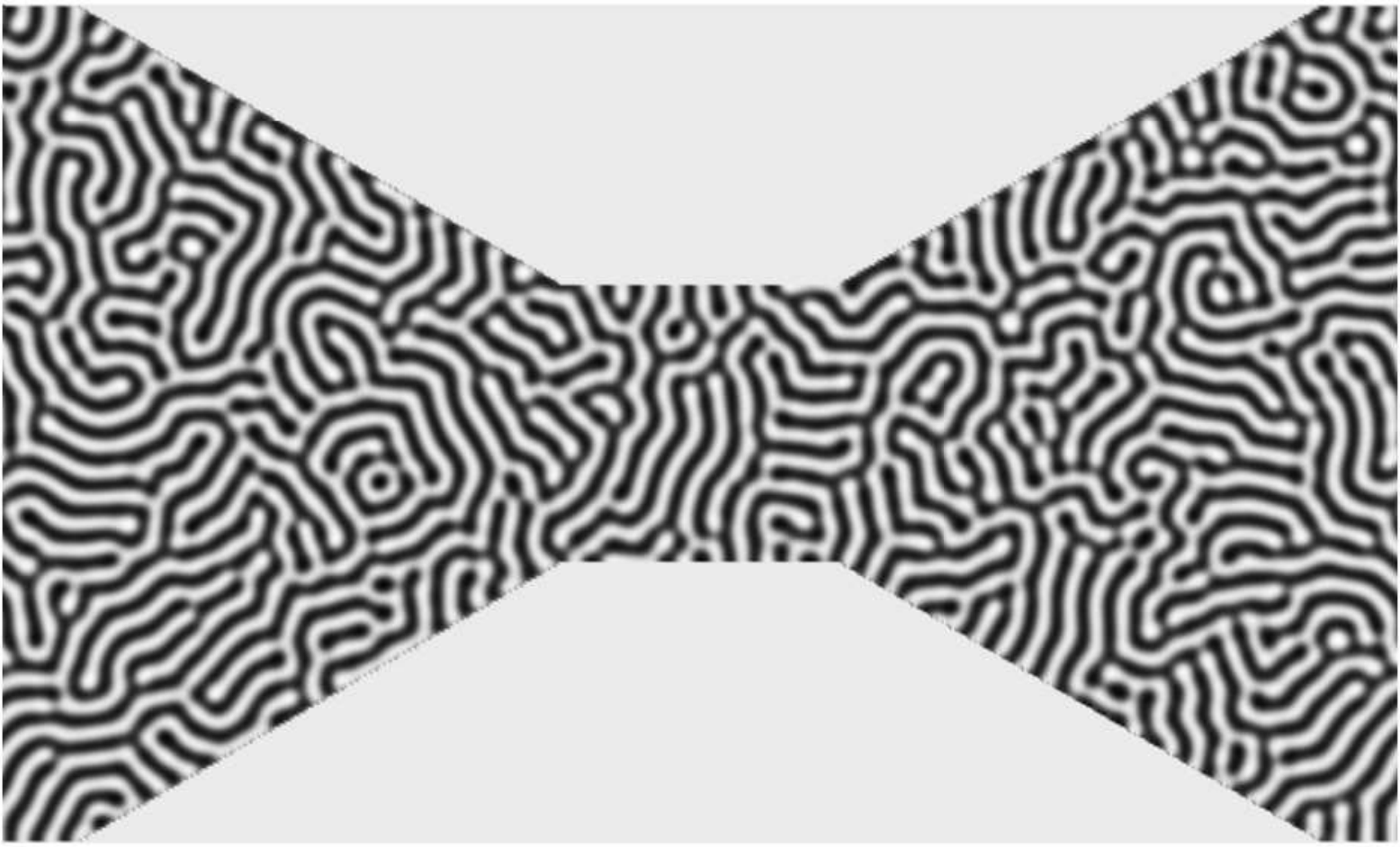}\\
$t=100 \Delta t$
\end{center}
\end{minipage}
\begin{minipage}{0.31\linewidth}
\begin{center}
\includegraphics[clip=true,width=1.6in]{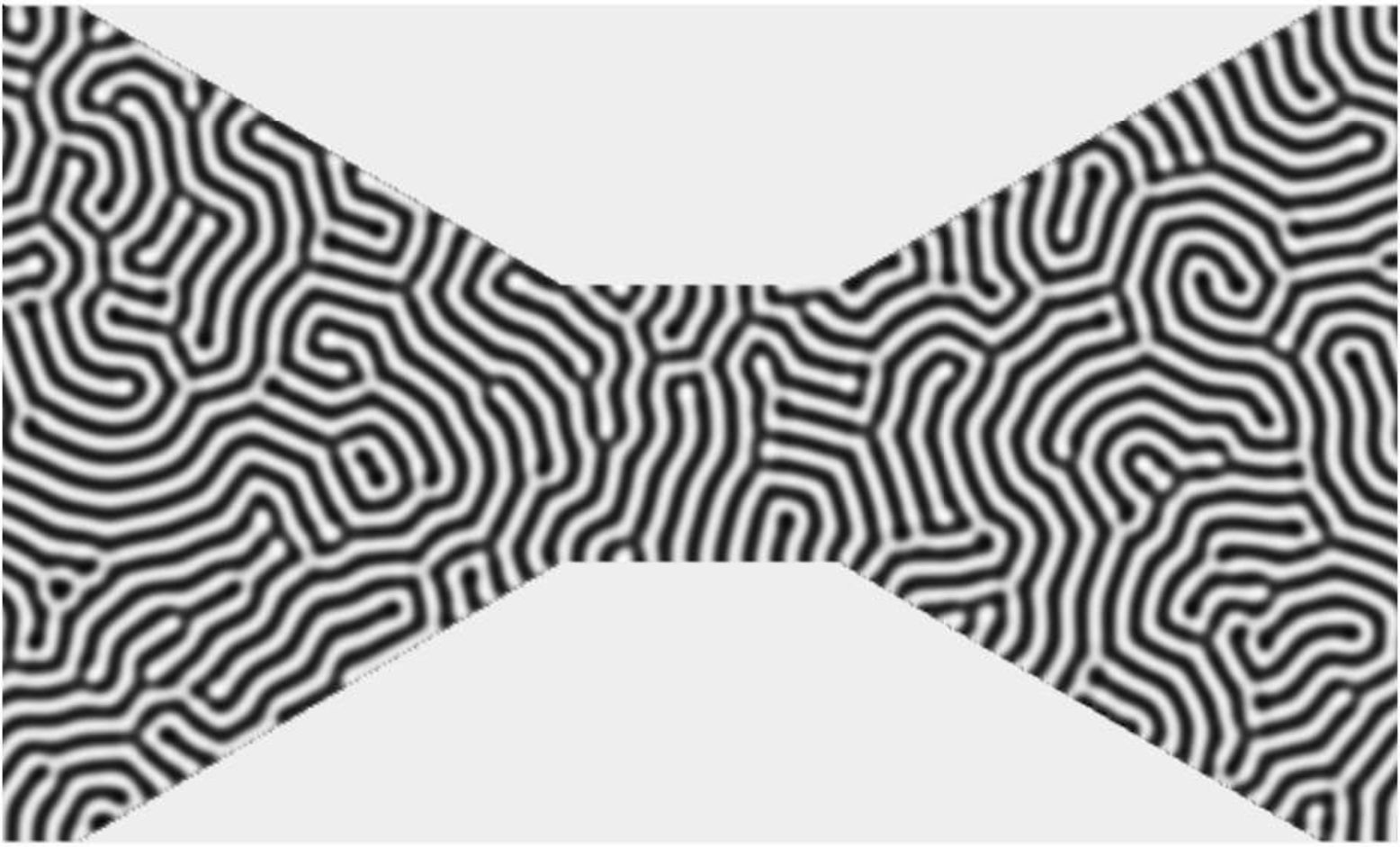}\\
$t=2000 \Delta t$
\end{center}
\end{minipage}\\
\caption{{Time evolution of $\phi$ when applying (a) the
Dirichlet and (b) the homogeneous Neumann boundary condition for
$\phi$ on $\Gamma_1$. The other boundary conditions are used to
be equal to the previous examples. Here, we denote the simulation
time on the bottom of figures columns line.}} \label{Compare_BC}
\end{figure}

In this section, we compare numerical results by the Dirichlet and
Neumann boundary conditions. We have the comparison test on the same
geometry shown in figure~\ref{angle}~(c). Figure~\ref{Compare_BC}~(a) shows the temporal evolution of $\phi$ when
applying Dirichlet and homogeneous Neumann conditions for $\phi$ and
$\mu$ on the boundary $\Gamma_1$, respectively. Figure~\ref{Compare_BC}~(b) represents the temporal evolution of $\phi$
when applying homogeneous Neumann condition for $\phi$ and $\mu$ on
the boundary $\Gamma_1$. As we expected, we obtain the lamella
pattern in the narrow channel when we apply the Dirichlet boundary
condition on $\Gamma_1$. However, the numerical solution with the
zero homogeneous boundary condition for $\phi$ has many defects in
the narrow channel and a contact angle of $90^\circ$ on all
boundaries.

\section{Conclusions} \label{Conc}

 In this paper, we numerically investigated the local
defectiveness control of self-assembled diblock copolymer patterns
through appropriate substrate design. We used a nonlocal
Cahn-Hilliard equation for the phase separation dynamics of
diblock copolymers. We discretized the nonlocal CH equation by an
unconditionally stable finite difference scheme on a tapered trench
design and, in particular, we used Dirichlet, Neumann, and
periodic boundary conditions. The value at the Dirichlet boundary is
obtained from energy-minimizing wavelength. We solved the resulting
discrete equations using the Gauss-Seidel iterative method. We performed
various numerical experiments to know the effect of the channel width,
length, and angle. Our simulation results were consistent with real
experimental observations.


\section*{Acknowledgement}

The first author (D.~Jeong) was supported by a Korea University
Grant. The corresponding author (J.S.~Kim) was supported by the
National Research Foundation of Korea (NRF) grant funded by the
Korea government (MSIP) (NRF-2014R1A2A2A01003683).



\ukrainianpart
\title{Числове дослідження керування локальною дефектністю структур діблок-кополімерів}
\author{Д. Йонг, Й. Чоі, Ю. Кім}
\address{Факультет математики, Корейський університет, Сеул 136-713, Республіка Корея}

\makeukrtitle

\begin{abstract}
Проведено числове  дослідження керування локальною дефектністю
самоорганізованих структур діблок-кополімерів за допомогою відповідної  конструкції субстрату.
Використовується нелокальне рівняння Кана-Хілларда  для динаміки фазового розділення діблок-кополімерів.
Здійснено дискретизацію нелокального рівняння з використанням  безумовно стійкої схеми скінченної різниці на звуженій канавці зразка
і, зокрема, використано крайові умови  Діріхле, Ньюмана і періодичні граничні умови.
Значення  при крайових умовах Діріхле отримано згідно з рівноважним ламеларним профілем, що відповідає енергетичному мінімуму.
Ми розв'язуємо отримані дискретні рівняння, використовуючи ітеративний метод Гаусса-Зейделя.
Проведено різні числові експерименти, такі як вплив ширини каналу, довжини каналу та кута на динаміку фазового розділення.
Результати симуляцій  відповідають попереднім експериментальним спостереженням.
\keywords діблок-кополімери, нелокальне рівняння Кана-Хілларда, керування локальною дефектністю
\end{abstract}

\end{document}